\documentclass[11pt]{article}
\pdfoutput=1
\usepackage[utf8]{inputenc}

\usepackage{jheppub}
\usepackage{bigints}

\usepackage{bbm}

\usepackage{bbold}

\usepackage{tabularx}

\usepackage{tikz}

 \usepackage{diagbox}

\usepackage{ae}
\usepackage[T1]{fontenc}
\usepackage{url}

\usepackage{tabularx}

\usepackage{graphicx}
\usepackage{dsfont}
\usepackage{setspace}
\usepackage{amsfonts,amsmath,amsthm,amssymb,amsbsy}
\usepackage{longtable}
\usepackage{array}
\usepackage{color}
\usepackage{needspace}
\usepackage[numbers]{natbib}
\usepackage{colortbl}

\usepackage{enumitem}

\usepackage{fancyhdr}
 \newcommand{\Le}{\textup{\protect\scalebox{-1}[1]{L}}}

\title{From Momentum Amplituhedron Boundaries to Amplitude Singularities and Back}

\author[1,2]{Livia Ferro,}\emailAdd{livia.ferro@lmu.de}
\author[2]{Tomasz \L ukowski,}\emailAdd{t.lukowski@herts.ac.uk}
\author[2]{Robert Moerman}\emailAdd{r.moerman@herts.ac.uk}
\affiliation[1]{Arnold--Sommerfeld--Center for Theoretical Physics,\\ Ludwig--Maximilians--Universit\"at, \\ Theresienstra\ss e 37, 80333 M\"unchen, Germany }
\affiliation[2]{School of Physics, Astronomy and Mathematics, \\ University of Hertfordshire, \\  Hatfield, Hertfordshire, AL10 9AB, United Kingdom}

\abstract{The momentum amplituhedron is a positive geometry encoding tree-level scattering amplitudes in  $\mathcal{N}=4$ super Yang-Mills directly in spinor-helicity space. In this paper we classify all boundaries of the momentum amplituhedron $\mathcal{M}_{n,k}$ and explain how these boundaries are related to the expected factorization channels, and soft and collinear limits of tree amplitudes. Conversely, all physical singularities of tree amplitudes are encoded in this boundary stratification. Finally, we find that the momentum amplituhedron $\mathcal{M}_{n,k}$ has Euler characteristic equal to one, which provides a first step towards proving that it is homeomorphic to a ball.}
\begin{document}
\begin{flushright}
{\small LMU-ASC 12/20}
\end{flushright}
\maketitle

%%%%%%%%%%%%%%%%%%%%%%%%%%%%%%%%%%%%%%%%
%%%%%%%%%%%%%%%%%%%%%%%%%%%%%%%%%%%%%%%%

\section{Introduction and Motivation}

Recent years have seen a surge in novel geometric constructions describing physical quantities in quantum field theories. In particular, it has been shown that scattering amplitudes in (planar) $\mathcal{N}=4$ super Yang-Mills theory (sYM) are encoded in geometric spaces called ``Amplituhedra".
These are examples of a recently introduced class of interesting geometries called {\it positive geometries} \cite{Arkani-Hamed:2017tmz}. The latter are regions with boundaries inside projective spaces, equipped with rational differential forms which have the defining property that they are logarithmic on approaching any boundary of the positive geometry. 
Amplituhedra form a family of positive geometries labelled by discrete parameters which, in the language of amplitudes, translate as the number of particles ($n$), the total helicity of the amplitude ($k$) and an additional parameter ($m$) which in the physical case takes the value $m=4$. We distinguish two types of amplituhedra depending on which space they are defined: the ordinary amplituhedron $\mathcal{A}_{n,k}^{(m)}$ \cite{Arkani-Hamed:2013jha} is defined on momentum twistor space \cite{Hodges:2009hk}, while the momentum amplituhedron $\mathcal{M}_{n,k}^{(m)}$ \cite{Damgaard:2019ztj}, for even $m$, is defined on the space of spinor-helicity variables. At the moment, we know much more about the ordinary amplituhedra, which for $k=1$ are familiar objects -- cyclic polytopes -- and for larger $k$ provide a generalization of convex polytopes into the Grassmannian space. Similar to polytopes, they have an intricate combinatorial structure of boundaries, which has already been classified for the first few examples: for $m=1$ in \cite{Karp:2016uax} and for $m=2$ in \cite{Lukowski:2019kqi}. Less is known about their topology and the only available result is the one in \cite{Galashin:2017onl}, where it was proven that the $m=1$ amplituhedron is homeomorphic to a $k$-dimensional ball. For the momentum amplituhedron even less is known: for $m=2$ it was shown in \cite{Lukowski:2020dpn} that $\mathcal{M}_{n,k}^{(2)}$ shares many properties with the hypersimplex $\Delta_{k+1,n}$, which is a well studied convex polytope with known boundary structure.
However, from the point of view of physics, we are mostly interested in the case when $m=4$ for which many features are yet to be discovered and proven. For instance, it is conjectured that amplituhedra encode physical singularities of scattering amplitudes in the structure of their boundaries. While this can be straightforwardly seen for the codimension one boundaries, which encode factorisation channels and collinear limits, a careful study of all boundaries deeper in the geometry is still missing.

In this paper we fill this gap by studying the boundary structure of the momentum amplituhedron $\mathcal{M}_{n,k}\equiv\mathcal{M}_{n,k}^{(4)}$. We show how the physical singularities of the amplitude are encoded in the boundaries of this geometry. Moreover, for all cases studied, we find their Euler characteristic to be one, strongly indicating that the momentum amplituhedron is a $(2n-4)$-dimensional ball. This is a remarkable fact, which shows an advantage in studying the momentum amplituhedron compared to the ordinary amplituhedron for which the structure of boundaries is much more complicated and as yet unknown.

This paper is organised as follows. In section \ref{sec:singularities} we review the singularities of scattering amplitudes, first for pure Yang-Mills theory and then for $\mathcal{N}=4$ sYM. We provide there an explicit form of the super-splitting functions and discuss in detail how to obtain soft limits as a combination of two consecutive collinear limits. In section \ref{sec:momamp} we review the definition of the momentum amplituhedron and explain how to find the complete stratification of its boundaries. We discuss the Euler characteristic of the momentum amplituhedron and construct the generating function encoding the number of its boundaries.

%%%%%%%%%%%%%%%%%%%%%%%%%%%%%%%%%%%%%%%
%%%%%%%%%%%%%%%%%%%%%%%%%%%%%%%%%%%%%%%%

%\section{Physics of Collinear and Soft Limits}

\section{Singularities of Scattering Amplitudes}
\label{sec:singularities}
Let us begin by reviewing the well-known behaviour of colour-ordered amplitudes under collinear and soft limits at tree level. These are governed by universal functions, which depend only on the particles which become collinear, or, in the case of soft limits, on the nearest neighbours of the soft particle. We first recall the known behaviour of collinear and soft gluons in pure Yang-Mills theory (see for instance \cite{Henn:2014yza} and references therein) in order to extend this analysis to $\mathcal{N}=4$ sYM. As a result, we provide an explicit form of the super-splitting functions for the case when two super-particles become collinear. One finds two such %collinear super-
functions, which correspond to a helicity-preserving and a helicity-decreasing case. For super-soft limits, we confirm that only the gluons present in the $\mathcal{N}=4$ sYM superfield have divergent behaviour when becoming soft. Again one finds two super-soft limits and each of them can be thought of as two simultaneous collinear limits of the same type. 

\subsection{Pure Yang-Mills Theory}
Colour-ordered amplitudes at tree- and loop-level exhibit known factorization properties when external momenta reach certain singular configurations. 
These limits provide both useful constraints for the evaluation of amplitudes and tests for the consistency  of obtained results. In what follows, we shall focus exclusively on tree-level amplitudes.
In this case, colour-ordered amplitudes can develop poles when the sum of adjacent momenta goes on shell: $P_{i,j}^2=(p_i+p_{i+1}+ \ldots +p_j)^2 \rightarrow 0$. On these poles, called  multi-particle  poles, amplitudes factorize  in the following way
\begin{equation}\label{factorization}
A_n^{\mathrm{tree}}(1,\ldots,n) \xrightarrow{\text{$P_{i,j}^2\rightarrow 0$}} \sum_{h= \pm1} ~A^{\mathrm{tree}}_{j-i+2}(i,\ldots,j, P_{i,j}^h) \frac{1}{P_{i,j}^2} ~ A^{\mathrm{tree}}_{n-j+i}(P_{i,j}^{-h},j+1, \ldots, i-1), 
\end{equation}
where the particles are cyclically ordered.

A special case of the above factorization formula (which we shall argue below) are collinear singularities. In general, collinear singularities arise when the momenta of two particles become proportional ($p_i \sim p_{i+1}$), which is equivalent to $P_{i,i+1}^2=0$. On the other hand, soft singularities arise when the on-shell momentum of one of the external particles goes to zero: $p_i\to0$. In these limits, scattering amplitudes at tree-level exhibit universal factorization properties given by
\begin{equation}
A_n^{\mathrm{tree}}(\ldots,i^{h_i}, (i+1)^{h_{i+1}}, \ldots) \xrightarrow{\text{$p_i || p_{i+1}$}} \sum_{h=\pm 1} ~A^{\mathrm{tree}}_{n-1}(\ldots, P_{i,i+1}^h, \ldots) ~ \mathrm{Split}^{\mathrm{tree}}_{-h}(i^{h_i}, (i+1)^{h_{i+1}}), 
\end{equation}
\begin{equation}
A_n^{\mathrm{tree}}(\ldots,i,s^h,j, \ldots) \xrightarrow{\text{$p_s\rightarrow 0$}} A^{\mathrm{tree}}_{n-1}(\ldots, i,j, \ldots) ~ \mathrm{Soft}^{\mathrm{tree}}(i,s^h,j),
\end{equation}
where $\mathrm{Split}^{\mathrm{tree}}$ and $\mathrm{Soft}^{\mathrm{tree}}$ are universal functions known as the tree-level splitting and soft functions, respectively. 

In order to argue that collinear limits arise as a particular case of the factorization formula \eqref{factorization}, let us, without loss of generality, take the particles with momenta $p_1$ and $p_2$ to be collinear. In this case, the momentum $P_{12}\equiv P_{1,2}=p_1+p_2$ goes on-shell and  \eqref{factorization} becomes
\begin{align}
A_n^\text{tree}(1,2,\dots,n)\xrightarrow{P_{12}^2\to 0}\sum_{h=\pm 1} A_3^\text{tree}(1,2,P_{12}^h)\frac{1}{P_{12}^2}A_{n-1}^\text{tree}(P_{12}^{-h},3,\dots,n),
\end{align}
where
\begin{align}
P_{12}^2=(p_1+p_2)^2=2 p_1\cdot p_2=\langle 12\rangle [12] \rightarrow 0 \,.\label{eq:col-momenta}
\end{align}
Note that when the brackets $[12]$ and $\langle 12\rangle$ are independent, then the collinearity condition $p_1\cdot p_2=0$ in \eqref{eq:col-momenta} can be achieved either by taking $\langle 12\rangle\to0$ or $[12]\to0$.
The independence of these two limits follows from three-particle special kinematics: for three-particle amplitudes to be non-vanishing and satisfy momentum conservation, angle and square brackets must be allowed to be independent. The independence of angle and square brackets can be accomplished by considering complex momenta (or $(2,2)$ space-time signature). As we shall show later for $\mathcal{N}=4$ sYM, $\langle 12\rangle\to0$ parametrises the helicity-preserving collinear limit, while $[12]\to0$ describes the collinear limit which decreases the helicity of the resulting amplitude by one.

To make a direct connection to the description of collinear limits in terms of splitting functions, let us examine $A_3^\text{tree}(1,2,P_{12}^h)\frac{1}{P_{12}^2}$ for a particular helicity assignment, say $(1^+,2^-,P_{12}^+)$. In this instance
\begin{align}
A_3^\text{tree}(1^+,2^-,P_{12}^+)\frac{1}{P_{12}^2}=\frac{[1P_{12}]^4}{[12][2P_{12}][P_{12}1]}\frac{1}{\langle 12\rangle[12]}=-\frac{[1P_{12}]^3}{[12]^2[1P_{12}]}\frac{1}{\langle 12\rangle}. \label{eq:col-ex-pmp}
\end{align}
From three-particle special kinematics for two positive-helicity gluons and one negative-helicity gluon, we know that all angle brackets vanish for $A_3^\text{tree}(1^+,2^-,P_{12}^+)$ and therefore $\langle 12\rangle\to0$ parametrizes the singularity in \eqref{eq:col-ex-pmp}. In particular, three-particle special kinematics requires
\begin{align}
\lambda_{P_{12}}=\alpha_1\lambda_1=\alpha_2\lambda_2, \label{eq:col-ex-pmp-lambda}
\end{align}
with
\begin{align}
\tilde\lambda_{P_{12}}=\alpha_1^{-1}\tilde\lambda_1+\alpha_2^{-1}\tilde\lambda_2 \,,
\label{eq:col-ex-pmp-lambdat}
\end{align}
following from momentum conservation. This allows us to rewrite the three-point amplitude as
\begin{align}
A_3^\text{tree}(1^+,2^-,P_{12}^+)\frac{1}{P_{12}^2}=\frac{\alpha_1}{\alpha_2^3}\frac{1}{\langle 12\rangle}. \label{eq:col-ex-pmp-alphas}
\end{align}
Having removed any dependence on square brackets, we can now restrict our attention to real external momenta which forces $\lambda$ and $\tilde\lambda$ to be conjugate to each other:
\begin{align}\label{parametrization.collinear}
\lambda_1&=\alpha_1^{-1}\lambda_{P_{12}}=\sqrt{z}\lambda_{P_{12}}, & \lambda_2&=\alpha_2^{-1}\lambda_{P_{12}}=\sqrt{1-z}\lambda_{P_{12}} ,\\
\tilde\lambda_1&=\alpha_1^{-1}\tilde\lambda_{P_{12}}=\sqrt{z}\tilde\lambda_{P_{12}}, & \tilde\lambda_2&=\alpha_2^{-1}\tilde\lambda_{P_{12}}=\sqrt{1-z}\tilde\lambda_{P_{12}},
\end{align}
where $z$ parametrizes the fraction of the total momentum $P_{12} = p_1 + p_2$ ($0 \leq z \leq 1$) carried by $p_1$ and $p_2$.
In this case, \eqref{eq:col-ex-pmp-alphas} reduces to the well-known splitting function
\begin{align}
A_3^\text{tree}(1^+,2^-,P_{12}^+)\frac{1}{P_{12}^2}=\text{Split}_{+}^{\text{tree}}(z;1^+,2^-)\equiv\frac{(1-z)^2}{\sqrt{z(1-z)}}\frac{1}{\langle 12\rangle} \,.
\label{eq:col-ex-pmp-zs}
\end{align}
Similarly, repeating the above analysis for different helicity assignments produces 
\begin{align}
A_3^\text{tree}(1^-,2^+,P_{12}^+)\frac{1}{P_{12}^2}&=\text{Split}_{+}^{\text{tree}}(z;1^-,2^+)= \frac{z^2}{\sqrt{z(1-z)}}\frac{1}{\langle 12\rangle},\\
A_3^\text{tree}(1^-,2^-,P_{12}^+)\frac{1}{P_{12}^2}&=\text{Split}_{+}^{\text{tree}}(z;1^-,2^-)=\frac{1}{\sqrt{z(1-z)}}\frac{1}{[12]},\\
A_3^\text{tree}(1^+,2^+,P_{12}^+)\frac{1}{P_{12}^2}&=\text{Split}_{+}^{\text{tree}}(z;1^+,2^+)= 0 \,.
\end{align}
The remaining splitting functions are obtained in the same way and are given by
\begin{align}
\text{Split}_{-}^{\text{tree}}(z;1^{h_1},2^{h_2})=\text{Split}_{+}^{\text{tree}}(z;1^{-h_1},2^{-h_2})\Big|_{[12]\leftrightarrow\langle 12\rangle}.
\end{align}
This analysis motivates our diagrammatic notation employed later on in this paper in which we indicate collinear limits by attaching three-point amplitudes.

Having demonstrated how collinear limits follow from factorization on two-particle poles, we now argue that taking consecutive collinear limits of the same type produces a soft limit\footnote{Consecutive collinear limits of different types do not lead to a divergent behaviour.}. To this end, let us consider the explicit example of $A_{n}(1^+,2^+,3,\ldots,n)$ (with unspecified helicity assignments for particles $3,\ldots,n$) and consider the two consecutive collinear limits $\langle 12\rangle\to0$ and $\langle 23\rangle\to 0$, keeping $\langle 13\rangle$ generic. When particles $1$ and $2$ become collinear we find that
\begin{equation}
A_n(1^+,2^+,3,\ldots,n)\to A_3(1^+,2^+,P_{12}^-)\frac{1}{P_{12}^2}A_{n-1}(P_{12}^+,3,\ldots,n),
\end{equation}
using the parametrisation given in \eqref{parametrization.collinear}. If we additionally take the limit $\langle 23\rangle \to 0$, we can parametrise this limit as
\begin{align}\label{parametrization.collinear2}
\lambda_2&=\sqrt{w}\lambda_{Q_{23}}, & \lambda_3&=\sqrt{1-w}\lambda_{Q_{23}},\\
\tilde\lambda_2&=\sqrt{w}\tilde\lambda_{Q_{23}}, & \tilde\lambda_3&=\sqrt{1-w}\tilde\lambda_{Q_{23}},
\end{align}
where now the momentum $Q_{23}= p_2 +p_3$ is distributed as $p_2 = w\, Q_{23}$ and $p_3 = (1-w) Q_{23}$. We obtain
\begin{equation}
A_3(1^+,2^+,P_{12}^-)\frac{1}{P_{12}^2}=\frac{1}{\sqrt{z(1-z)}}\frac{1}{\langle 12\rangle} \xrightarrow{p_2 || p_3} \frac{1}{z}\frac{\langle 13\rangle }{\langle 12\rangle\langle 23\rangle},
\end{equation}
where the dependence on the parameter $w$ drops out entirely. We want to guarantee that the momenta $p_1$ and $p_3$ are independent. The only way to achieve this is by taking the limit $z\to 1$ and $w\to 0$ which implies that $P_{12}=p_1$ and $Q_{23}=p_3$ giving
\begin{equation}
A_n(1^+,2^+,3,\ldots,n)\to \frac{\langle 13\rangle}{\langle 12\rangle\langle 23\rangle} A_{n-1}(1^+,3,\ldots,n).
\end{equation}
A similar calculation can be done for other helicity configurations. One finds that the soft behaviour of an amplitude depends only on the helicity of the soft particle and on the momenta of its two closest neighbours, but not on their helicities. For a soft positive helicity gluon, the soft factor is given by
\begin{equation}
 \mathrm{Soft}^{\mathrm{tree}}(i,s^+,j)=\frac{\langle ij\rangle}{\langle is\rangle \langle sj\rangle},
 \end{equation}
 while for a soft negative helicity gluon  
 \begin{equation}
 \mathrm{Soft}^{\mathrm{tree}}(i,s^-,j)= \frac{[ij]}{[i s][s j]}.
 \end{equation}
 Here $(i,s,j)$ are three consecutive particles.

\subsection{\texorpdfstring{$\mathcal{N}=4$}{} Supersymmetric Yang-Mills Theory}
Let us start by recalling that the on-shell multiplet of $\mathcal{N}=4$ sYM can be collected into a single on-shell chiral superfield by means of the Grassmann-odd variables $\eta_A$ with $A=1,\dots,4$:
\begin{equation}\label{superfield}
\Phi=g^+ +\eta_A \, \lambda^A +\frac{1}{2!}\eta_A\eta_B \, S^{AB} +\frac{1}{3!}\eta_A\eta_B\eta_C \, \epsilon^{ABCD} \bar{\lambda}_{D}+\frac{1}{4!}\eta_A\eta_B\eta_C\eta_D \epsilon^{ABCD}\, g^-,
\end{equation}
with gluons $g^+$ and $g^-$, gluinos  $\lambda^A$ and $\bar{\lambda}_{D}$, and scalars $S^{AB}$.
A generic $n$-particle superamplitude $\mathcal{A}_n = \mathcal{A}_n(\Phi_1,\Phi_2,\ldots,\Phi_n)$ can be expanded in terms of helicity sectors
\begin{align}
\mathcal{A}_n&=A_{n,2} + A_{n,3} + \cdots + A_{n,n-2}, & &(n\ge 4)
\end{align}
where $A_{n,k}$ is the superamplitude for the N$^{k-2}$MHV sector and has Grassmann degree $4k$. Superamplitudes  factorize on multi-particle poles in a very similar way as for pure Yang-Mills theory
\begin{align}
A_{n,k}(1,\dots,n)&\xrightarrow{P_{i,j}^2\to 0}\\\nonumber
&\sum_{k'=1}^{k}\int d^4\eta_{P_{i,j}} A_{j-i+2,k'}(i,\ldots,j,P_{i,j})\frac{1}{P_{i,j}^2}A_{n-j+i,k-k'+1}(P_{i,j},j+1,\dots,i-1) \,.
\end{align}

 In the following we present a generalization of the splitting and soft functions which were discussed in the previous section. Let us start by considering the collinear limit. 
In the limit $P_{12}^2\to0$, $A_{n,k}$ factorizes as follows
\begin{align}\label{super.collinear}
A_{n,k}(1,2,3,\dots,n)\xrightarrow{P_{12}^2\to 0}\sum_{k'=1}^{2}\int d^4\eta_{P_{12}} A_{3,k'}(1,2,P_{12})\frac{1}{P_{12}^2}A_{n-1,k-k'+1}(P_{12},3,\dots,n) \,,
\end{align}
where the sum runs over two contributions: one where the helicity of the original superamplitude is preserved ($k'=1$) and one where the helicity is reduced by one ($k'=2$). The integration over $\eta_{P_{12}}$ corresponds to summing over all possible types of fields exchanged in this factorization channel.

Let us  investigate the helicity-decreasing contribution first. Following the steps for pure Yang-Mills theory, we consider
\begin{align}
A_{3,2}(1,2,P_{12})\frac{1}{P_{12}^2} = \frac{\delta^{(8)}(Q)}{\langle 12\rangle\langle 2P_{12}\rangle\langle P_{12}1\rangle}\frac{1}{\langle 12\rangle[12]} \,,
\label{eq:col-i2}
\end{align}
where $Q^{\alpha A} = \sum_{i=1}^n \lambda^{\alpha}_i \eta^A_i$.
With the three-particle special kinematics in mind, we take the limit $[12]\rightarrow 0$ and find
\begin{align}
A_{3,2}(1,2,P_{12})\frac{1}{P_{12}^2} = \frac{\delta^{(8)}(Q)}{\langle 12\rangle^4}\frac{\alpha_1\alpha_2}{[12]} \,,
\end{align}
where we used $\lambda_{P_{12}}=\alpha_1^{-1}\lambda_1+\alpha_2^{-1}\lambda_2$.
We can further simplify the super-momentum-conserving delta function to obtain  
\begin{align}
A_{3,2}(1,2,P_{12})\frac{1}{P_{12}^2} =
\frac{\alpha_1\alpha_2}{[12]}\prod_{A=1}^{4}(\eta_{1A}\eta_{2A}+\alpha_2^{-1}\eta_{1A}\eta_{P_{12}A}-\alpha_1^{-1}\eta_{2A}\eta_{P_{12}A}).
\end{align}
Restricting to real momenta, we find that $\alpha_1$ and $\alpha_2$ are no longer independent and we can again use \eqref{parametrization.collinear} to parametrise this dependence and produce
\begin{align}
A_{3,2}(1,2,P_{12})\frac{1}{P_{12}^2} =
\text{Split}^\text{tree}_{-1}(z;\eta_1,\eta_2,\eta_{P_{12}}),
\end{align}
where
\begin{align}
\text{Split}^\text{tree}_{-1}(z;\eta_1,\eta_2,\eta_3)\equiv \frac{1}{\sqrt{z(1-z)}}\frac{1}{[12]}\prod_{A=1}^{4}(\eta_{1A}\eta_{2A}+\sqrt{1-z}\eta_{1A}\eta_{3A}-\sqrt{z}\eta_{2A}\eta_{3A})\label{eq:col-split-m}
\end{align}
is the \emph{helicity-decreasing tree-level super-splitting function}. One immediately recovers the pure Yang-Mills splitting functions by expanding \eqref{eq:col-split-m} and focusing on terms proportional to $\eta_1^4 \eta_2^4$, etc.

\begin{figure}
\begin{center}
\begin{tabular}{cc}
\includegraphics[scale=0.3]{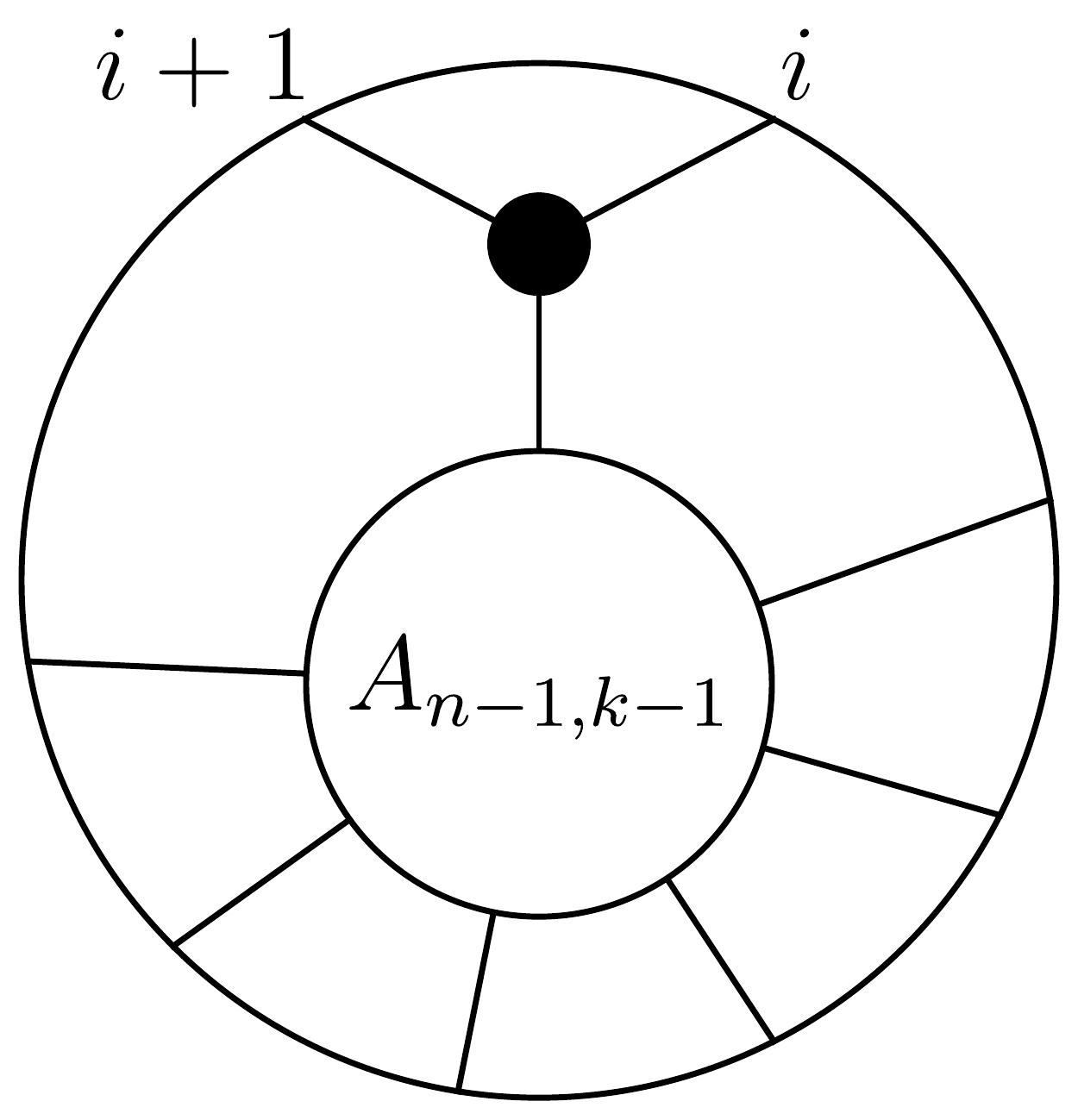}&\includegraphics[scale=0.3]{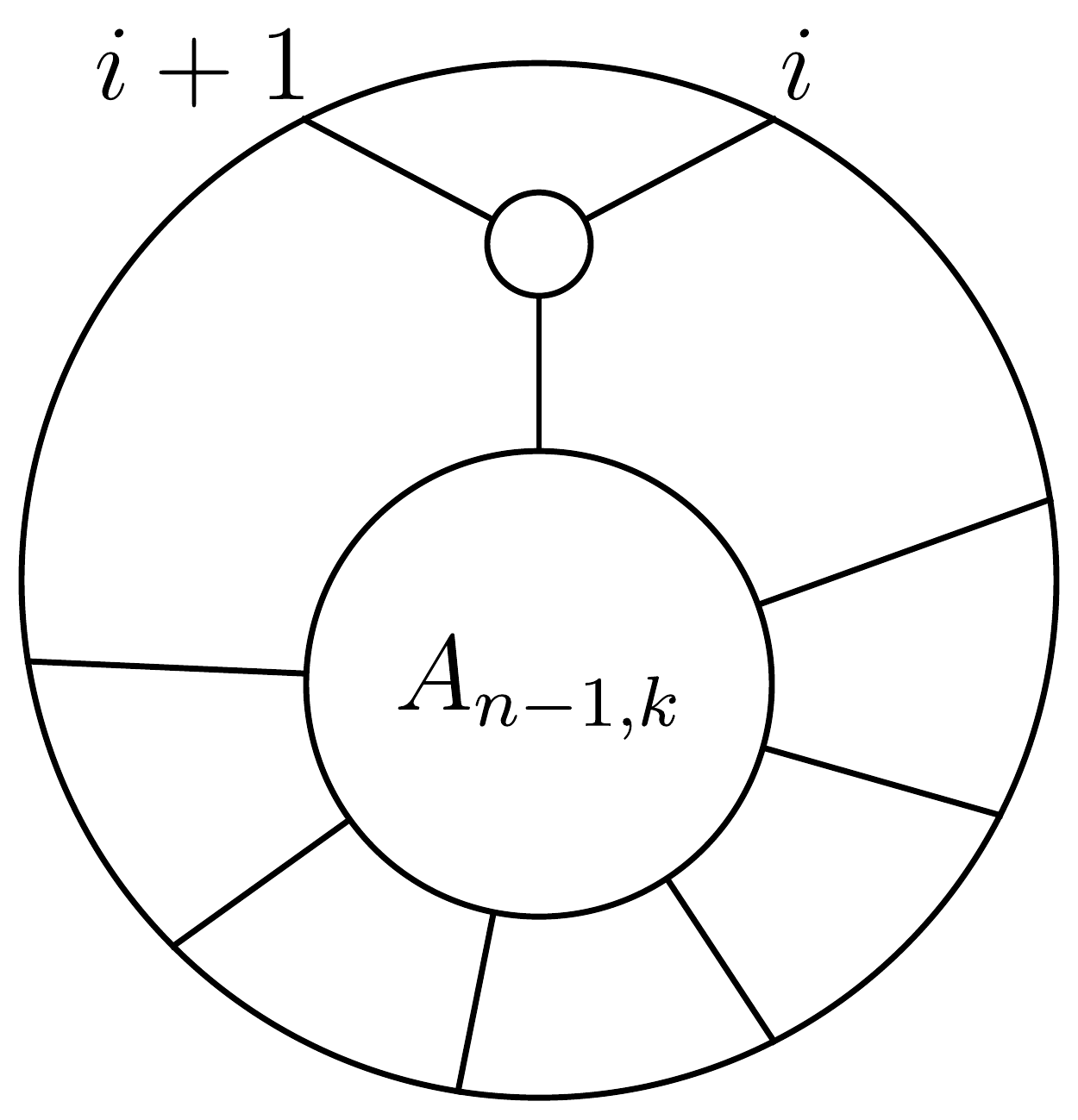}\\
(a)~$ [ ii+1] = 0$&
(b) $ \langle ii+1\rangle = 0$
\end{tabular}
\caption{Representation of collinear limits. In Figure 1(a) we represent the super-splitting function $\text{Split}^\text{tree}_{-1}$ with a black trivalent vertex. In Figure 1(b), $\text{Split}^\text{tree}_{0}$ is represented by a white trivalent vertex.}
\label{fig:collinear_sym}
\end{center}
\end{figure}

Repeating the above analysis for the helicity-preserving case, one can show that in the collinear limit $\langle 12 \rangle \rightarrow 0$ we have
\begin{align}
A_{3,1}(1,2,P_{12})\frac{1}{P_{12}^2} =
\text{Split}^\text{tree}_{0}(z;\eta_1,\eta_2,\eta_{P_{12}}),
\end{align}
where 
\begin{align}
\text{Split}^\text{tree}_{0}(z;\eta_1,\eta_2,\eta_3)\equiv \frac{1}{\sqrt{z(1-z)}}\frac{1}{\langle 12\rangle}\prod_{A=1}^{4}(\eta_{3A}-\sqrt{z}\eta_{1A}-\sqrt{1-z}\eta_{2A}) 
\label{eq:col-split-0}
\end{align}
is the \emph{helicity-preserving tree-level super-splitting function}. Again, one can easily show that the pure Yang-Mills theory splitting functions can be obtained by expanding \eqref{eq:col-split-0} and focusing on terms proportional to $\eta_1^4$, etc.
We observe from \eqref{eq:col-split-0} and \eqref{eq:col-split-m} that helicity-preserving and helicity-decreasing collinear limits are parametrised by $\langle 12\rangle\to0$ and $[12]\to0$, respectively, and that these are two physically distinguishable processes, which we depict in Fig.~\ref{fig:collinear_sym}. Although we believe these to be known results, to the best of our knowledge, the expressions for the super-splitting functions given in \eqref{eq:col-split-0} and \eqref{eq:col-split-m} do not appear in the literature, and we therefore record them here for convenience.  

\begin{figure}
\begin{center}
\begin{tabular}{ccc}
\includegraphics[scale=0.3]{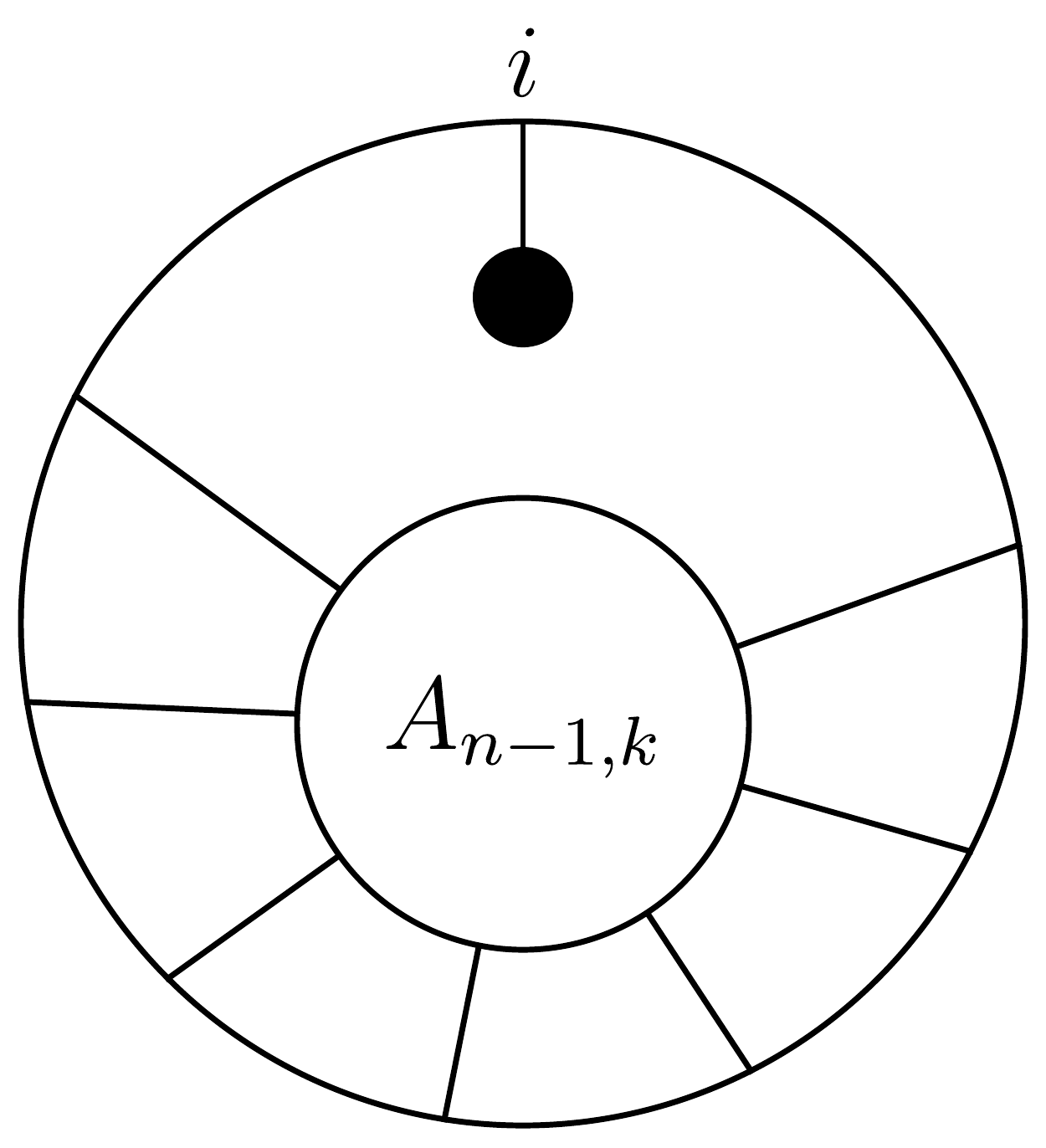}&\quad&\includegraphics[scale=0.3]{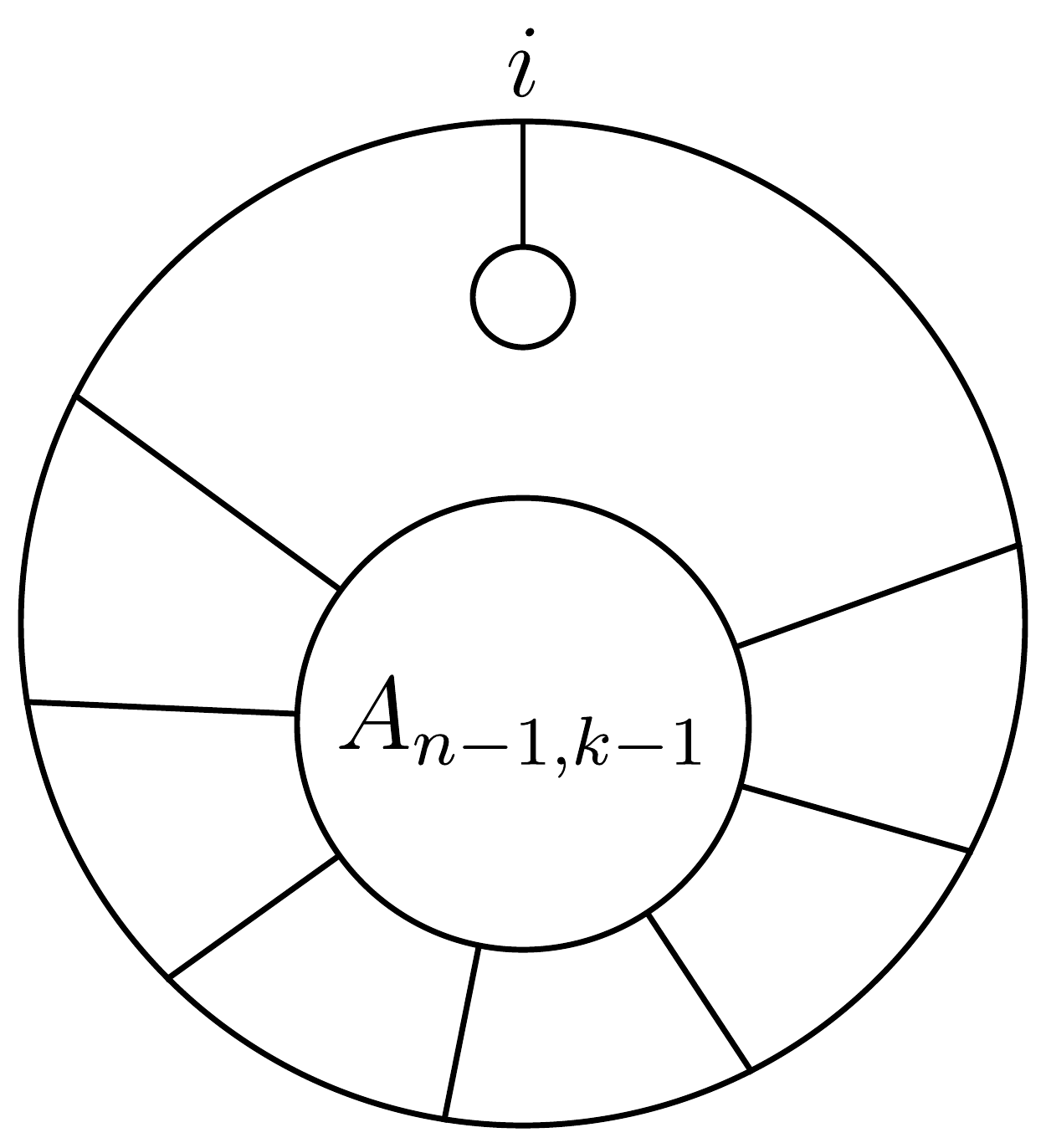}\\
(a) ~ $\langle ii+1\rangle=0=\langle i-1i\rangle$&&
(b) ~ $[ii+1]=0=[i-1i]$
\end{tabular}
\caption{Soft limits of an amplitude with the soft particle indicated by a black (white) lollipop. These correspond to two consecutive collinear limits of angle (square) brackets.}
\label{fig:soft_sym}
\end{center}
\end{figure}

As for pure Yang-Mills theory, we can also take consecutive collinear limits to produce a soft limit. Let us start with the helicity-preserving case. Using the parametrizations \eqref{parametrization.collinear} and \eqref{parametrization.collinear2}, and taking the limit $z\to 1$, we obtain
\begin{align}
\text{Split}^\text{tree}_{0}(z;\eta_1,\eta_2,\eta_{P_{12}})\to \frac{\langle 13\rangle}{\langle 12\rangle\langle 23 \rangle}\prod_{A=1}^{4}(\eta_{P_{12}A}-\eta_{1A}).
\end{align}
Substituting this into \eqref{super.collinear} and integrating over $\eta_{P_{12}}$ we find the \emph{helicity-preserving soft limit}
\begin{equation}\label{eq.soft.preserving.super}
A_{n,k}(1,2,3,\ldots,n)\to \frac{\langle 13\rangle}{\langle 12\rangle \langle 23\rangle} A_{n-1,k}(1,3,\ldots,n).
\end{equation}
Similarly, for consecutive helicity-reducing collinear limits we find
\begin{equation}
\text{Split}^\text{tree}_{-1}(z;\eta_1,\eta_2,\eta_{P_{12}})\to \frac{[13]}{[ 12][ 23 ]}\prod_{A=1}^{4}\eta_{2A}(\eta_{P_{12}A}-\eta_{1A}).
\end{equation}
and the \emph{helicity-decreasing soft limit} given by
\begin{equation}\label{eq.soft.decreasing.super}
A_{n,k}(1,2,3,\ldots,n)\to \eta_2^4\frac{[13]}{[ 12][ 23 ]} A_{n-1,k-1}(1,3,\ldots,n).
\end{equation}
Notice that for soft limits the only divergent contribution comes from soft gluons: in the first case, only the positive-helicity gluon contributed to  \eqref{eq.soft.preserving.super}, while in the second case, \eqref{eq.soft.decreasing.super} comes solely from the negative-helicity gluon. We depict the soft limits in Fig.~\ref{fig:soft_sym}.

Until now, we have considered $\mathcal{N}=4$ sYM written in the chiral superspace for which the superfield is given by \eqref{superfield}, i.e.\ $(\lambda^{\alpha},\tilde\lambda^{\dot\alpha},\eta^A)$. In order to connect our considerations with the following section, we need to rewrite all formulae in the non-chiral superspace, which consists of two $\eta$'s and two $\tilde\eta$'s, i.e.\ $(\lambda^{\alpha},\eta^{\alpha}|\tilde\lambda^{\dot\alpha},\tilde\eta^{\dot\alpha})$, where the latter are the Fourier transform of two of the $\eta$'s. We can accomplish this by performing Fourier transforms of all our formulae with respect to $\eta_3$ and $\eta_4$ which will introduce their conjugate fermionic coordinates $\tilde\eta_1$ and $\tilde\eta_2$. The main conclusions from our calculations will remain unchanged and we get the following formulae for the super-splitting functions
\begin{eqnarray*}
\text{Split}^\text{tree}_{-1}(z;\eta_1,\eta_2,\eta_3,\tilde\eta_1,\tilde\eta_2,\tilde\eta_3)&\to& \frac{1}{\sqrt{z(1-z)}}\frac{1}{[12]}\prod_{\alpha=1}^{2}(\eta_{1\alpha}\eta_{2\alpha}+\sqrt{1-z}\eta_{1\alpha}\eta_{3\alpha}-\sqrt{z}\eta_{2\alpha}\eta_{3\alpha}) \\
&&\times\prod_{\dot\alpha=1}^{2}(\tilde\eta_{3\dot\alpha}-\sqrt{z}\tilde\eta_{1\dot\alpha}-\sqrt{1-z}\tilde\eta_{2\dot\alpha}) \,, \\
\text{Split}^\text{tree}_{0}(z;\eta_1,\eta_2,\eta_3,\tilde\eta_1,\tilde\eta_2,\tilde\eta_3)&\to& \frac{1}{\sqrt{z(1-z)}}\frac{1}{\langle 12\rangle}\prod_{\alpha=1}^{2}(\eta_{3\alpha}-\sqrt{z}\eta_{1\alpha}-\sqrt{1-z}\eta_{2\alpha})\\
&&\times\prod_{\dot\alpha=1}^{2}(\tilde\eta_{1\dot\alpha}\tilde\eta_{2\dot\alpha}+\sqrt{1-z}\tilde\eta_{1\dot\alpha}\tilde\eta_{3\dot\alpha}-\sqrt{z}\tilde\eta_{2\dot\alpha}\tilde\eta_{3\dot\alpha}) \,,
\end{eqnarray*}
and soft limits:
\begin{eqnarray*}
A_{n,k}(1,2,3,\ldots,n)&\xrightarrow{\langle12 \rangle\to 0,\langle 23\rangle\to0}\prod_{\dot\alpha=1}^2\tilde\eta_{2\dot \alpha}\, \frac{\langle 13\rangle}{\langle 12\rangle \langle 23\rangle} A_{n-1,k}(1,3,\ldots,n) \,,\\ 
A_{n,k}(1,2,3,\ldots,n)&\xrightarrow{[12 ]\to 0,[ 23]\to0}\prod_{\alpha=1}^2\eta_{2 \alpha}\, \frac{[13]}{[12][23]} A_{n-1,k-1}(1,3,\ldots,n) \,.
\end{eqnarray*}

%%%%%%%%%%%%%%%%%%%%%%%%%%%%%%%%%%%%%%%
%%%%%%%%%%%%%%%%%%%%%%%%%%%%%%%%%%%%%%
\section{Physical Singularities from the Momentum Amplituhedron}
\label{sec:momamp}
After having reviewed the singularities of tree-level scattering amplitudes in the previous section, we now want to describe how these are encoded in positive geometries. It is conjectured that the physical singularities of tree amplitudes correspond to the boundaries of the momentum amplituhedron. While for facets, i.e.\ codimension one boundaries, this has already been established \cite{Damgaard:2019ztj}, the lower dimensional boundaries also encode further multi-particle factorizations and multi-particle collinear limits which are present deeper in the positive geometry and have not been classified before.
We will start with a review of the momentum amplituhedron construction followed by a description of its boundaries and an algorithm for finding them. We also provide evidence that, for the extensive number of cases analysed, this positive geometry has Euler characteristic one which suggests that the momentum amplituhedron is homeomorphic to a ball and therefore it is simpler than the ordinary amplituhedron.

\subsection{Definition of Momentum Amplituhedron}
The tree momentum amplituhedron $\mathcal{M}_{n,k}^{(m)}$ has been introduced in \cite{Damgaard:2019ztj} for $m=4$ and generalized to any even $m$ in \cite{Lukowski:2020dpn}. We focus here only on the physical case $m=4$, denoted simply $\mathcal{M}_{n,k}$, which is the positive geometry encoding tree-level scattering amplitudes in $\mathcal{N}=4$ sYM directly in spinor-helicity space. The momentum amplituhedron $\mathcal{M}_{n,k}$ is defined as the image of the positive Grassmannian $G_+(k,n)$, that is a subset of the Grassmannian $G(k,n)$ consisting of elements described by matrices with all ordered maximal minors non-negative, through the map
\begin{equation}
\label{Phi}
\Phi_{(\Lambda,\tilde\Lambda)}:G_+(k,n)\to G(k,k+2)\times G(n-k,n-k+2)\,.
\end{equation}
Here  $\Lambda$ and $\tilde\Lambda$ are bosonised versions of the non-chiral superspace coordinates  $(\lambda^{\alpha},\eta^{\alpha}|\tilde\lambda^{\dot\alpha},\tilde\eta^{\dot\alpha})$. In particular, $\tilde \Lambda$ is a $(k+2)\times n$ positive matrix and $\Lambda$ is a $(n-k+2)\times n$ twisted matrix (see \cite{Galashin:2018fri} for the precise definition).
To each element $C=\{c_{\dot\alpha i}\}$ of the positive Grassmannian $G_{+}(k,n)$ the map $\Phi_{\Lambda,\tilde\Lambda}$ associates a pair of Grassmannian elements $(\tilde Y,Y)\in G(k,k+2)\times G(n-k,n-k+2)$ in the following way
\begin{align}
\tilde Y^{\dot{A}}_{\dot{\alpha}}=c_{\dot\alpha i}\,\tilde\Lambda_i^{\dot{A}}\,,\qquad Y^A_\alpha=c^\perp_{\alpha i}\,\Lambda_i^A\,,
\label{Y}
\end{align}
where $C^\perp=\{c^\perp_{\alpha i}\}$ is the orthogonal complement of $C$. 
Although the dimension of the $(Y,\tilde Y)$-space is $2k+2(n-k)=2n$, the image of the positive Grassmannian through the $\Phi_{(\Lambda,\tilde\Lambda)}$ map is $2(n-2)$-dimensional since one can show that the following relation holds true
\begin{equation}
(Y^\perp\Lambda)\cdot(\tilde Y^\perp \tilde\Lambda)=0,
\end{equation}  
and therefore the image is embedded in a surface of codimension four.

The boundary structure of the momentum amplituhedron $\mathcal{M}_{n,k}$ is closely related to the boundary structure of the positive Grassmannian $G_+(k,n)$ and, in particular, each boundary of $\mathcal{M}_{n,k}$ can be labelled by a subset of labels for $G_+(k,n)$. The positive Grassmannian has been studied by Postnikov \cite{Postnikov:2006kva} and is known to have a very rich and interesting combinatorial structure. Each boundary stratum of $G_+(k,n)$ is called a \emph{positroid cell} and can be labelled by a variety of combinatorial objects, including affine permutations, (equivalence classes of) plabic diagrams and $\Le$-diagrams. An {\it affine permutation} is a generalization of the ordinary permutation which allows for two types of fixed-points. It is a map $\pi:\{1,2,\ldots,n\}\to\{1,2,\ldots,2n\}$ such that $a\leq \pi(a)\leq a+n$ and it reduces to an ordinary permutation mod $n$. We will use affine permutations to label positroid cells $S_\pi\subset G_+(k,n)$, but also to label images of $S_\pi$ in the momentum amplituhedron. For a given positroid cell $S_\pi\subset G_+(k,n)$ we denote by $\dim_C \pi$ its dimension and by $\partial_C\pi$ its boundary stratification. We also denote by $\partial^{-1}_C\pi$ the inverse boundary stratification of $\pi$, i.e.\ the set of all positroid cells $S_{\pi'}\subset G_+(k,n)$ for which $\pi\in\partial_C\pi'$.

For each positroid cell $S_\pi\subset G_+(k,n)$ we define the momentum amplituhedron dimension $\dim_M\pi$ as
\begin{equation}
\dim_M\pi=\dim \Phi_{(\Lambda,\tilde\Lambda)}(S_\pi).
\end{equation}
It is always true that $\dim_C(\pi)\geq \dim_M(\pi)$ and we can distinguish two cases:
\begin{itemize}
\item $\dim_C(\pi)=\dim_M(\pi)$: we will refer to cells satisfying this condition as simplicial-like,
\item $\dim_C(\pi)>\dim_M(\pi)$: these cells are polytopal-like.
\end{itemize}
This distinction refers to properties of polytopes: a simplex is a polytope which cannot be subdivided into smaller polytopes without introducing new vertices. Similarly here, the simplicial-like cells are those for which their images cannot be subdivided into smaller images of positroid cells. This is not the case for polytopal-like cells. In particular, given a positroid cell $\pi$ for which $\dim_C(\pi)>\dim_M(\pi)$, we can find a collection of cells in its boundary stratification $\partial_C\pi$, with the same amplituhedron dimension as $\pi$. Moreover, there exists a (non-unique) subset $\{\pi_1,\ldots,\pi_r\}\in\partial_C\pi$ such that the images $\{\Phi_{(\Lambda,\tilde\Lambda)}(\pi_1),\ldots,\Phi_{(\Lambda,\tilde\Lambda)}(\pi_r)\}$ triangulate the image $\Phi_{(\Lambda,\tilde\Lambda)}(\pi)$. 

After this basic introduction to the momentum amplituhedron, we are ready to explore its boundary stratification. In particular, the facets of the momentum amplituhedron $\mathcal{M}_{n,k}$ have been studied in \cite{Damgaard:2019ztj} and they belong to one of the following classes:
\begin{equation}
\langle Y i i+1\rangle=0\,,\qquad [\tilde Y ii+1]=0\,, \qquad S_{i,i+1,\ldots,j}=0\,,
\end{equation}
where 
\begin{equation}
S_{i,i+1,\ldots ,j}=\sum\limits_{a<b=i}^j \langle Y a b\rangle [\tilde Y ab]\,
\end{equation}
are equivalent to the Mandelstam invariants written in the momentum amplituhedron space. The invariant brackets here are defined as
\begin{align}
\langle Y i j\rangle&=\epsilon_{A_1A_2\ldots A_{n-k+2}}  Y_1^{A_1} Y_2^{A_2} \ldots Y_{n-k}^{A_{n-k}} \Lambda_{i}^{A_{n-k+1}}\Lambda_{j}^{A_{n-k+2}}\,, \\
[\tilde{Y} i j]&=\epsilon_{\dot A_1\dot A_2\ldots \dot A_{k+2}} \tilde{Y}_1^{\dot A_1} \tilde{Y}_2^{\dot A_2} \ldots \tilde{Y}_{k}^{\dot A_{k}} \tilde\Lambda_{i}^{\dot A_{k+1}}\tilde\Lambda_{j}^{\dot A_{k+2}}\,.
\end{align}

In what follows, we would like to extend our understanding of the momentum amplituhedron boundary structure beyond the facets. To this end, we will follow the method used in \cite{Lukowski:2019kqi}: assume that we have found all momentum amplituhedron boundaries of momentum amplituhedron dimension larger than $d$. Let us study all positroid cells $\pi$ with momentum amplituhedron dimension $\dim_M \pi = d$. For a given cell $\pi$, there are two options:
\begin{itemize}
\item either the momentum amplituhedron dimensions for all inverse boundaries of $\pi$ are higher than the momentum amplituhedron dimension of $\pi$: 
$$\forall_{\pi'\in\partial^{-1}_C\pi}: \dim_M\pi'>\dim_M\pi \,,$$
\item or we can find a cell among the inverse boundaries of $\pi$ which has a higher Grassmannian dimension but the same momentum amplituhedron dimension as $\pi$: 
$$\exists_{\pi'\in\partial^{-1}_C\pi}:\dim_M\pi'=\dim_M\pi\text{ and }\dim_C\pi'>\dim_C\pi \,.$$
\end{itemize}
We only keep the former cells since the latter are necessarily elements of a triangulation of a boundary of the momentum amplituhedron. After discarding these latter cells, there is still a possibility that some of the remaining cell images are spurious boundaries, which arise as internal boundaries in triangulations of polytopal-like boundaries.
Spurious boundaries can be identified (and removed) because they belong to a single $(d+1)$-dimensional momentum amplituhedron boundary, while external boundaries belong to at least two such boundaries. This procedure allows us to find all external boundaries of dimension $d$. We can follow this procedure recursively, starting from the known codimension-one boundaries, and work our way down to zero-dimensional boundaries: points.

%%%%%%%%%%%%%%%%%%%%%%%%%%%%%%%

\subsection{Boundary Stratification}
The algorithm described above has been implemented in the Mathematica package \texttt{amplituhedronBoundaries} \cite{Lukowski:2020bya}. Using it we were able to find the momentum amplituhedron boundary stratifications for up to $n=9$ and for all $k$.  These results are summarised in Table~\ref{tab:fvectors}. A careful study of these stratifications lead us to postulate a dual graph notation, see Section~\ref{sec:dual}, which allowed us to conjecture the general form of momentum amplituhedron boundaries and extend our analysis up to $n=11$, see Table~\ref{table:generating.high}. In particular, we were able to calculate how many boundaries of a given dimension there are in $\mathcal{M}_{n,k}$ for up to $n=11$. In all these cases, we found that the Euler characteristic of the momentum amplituhedron equals one.

\begin{table}
\begin{center}
\begin{tabular}{c|ccccccccccccc}
$(n,k)\backslash ~d$&0&1&2&3&4&5&6&7&8&9&10&11&12\\
\hline
$(4,2)~~~~$&6&12&10&4&1\\
$(5,2)~~~~$&10&30&40&30&15&5&1\\
$(6,2)~~~~$&15&60&110&120&90&50&21&6&1\\
$(6,3)~~~~$&20&90&180&215&180&114&54&15&1\\
$(7,2)~~~~$&21&105&245&350&350&266&161&77&28&7&1\\
$(7,3)~~~~$&35&210&560&910&1050&938&665&350&119&21&1\\
$(8,2)~~~~$&28&168&476&840&1050&1008&784&504&266&112&36&8&1\\
$(8,3)~~~~$&56&420&1400&2870&4200&4788&4424&3262&1820&720&188&28&1\\
$(8,4)~~~~$&70&560&1960&4200&6426&7672&7420&5696&3264&1280&300&32&1
\end{tabular}
\caption{The number of boundaries of $\mathcal{M}_{n,k}$ of a given dimension.}
\label{tab:fvectors}
\end{center}
\end{table}

\begin{figure}
\begin{center}
\begin{tabular}{ccc}
\includegraphics[scale=0.3]{images/collinearwhite}&\includegraphics[scale=0.3]{images/collinearblack}&\includegraphics[scale=0.3]{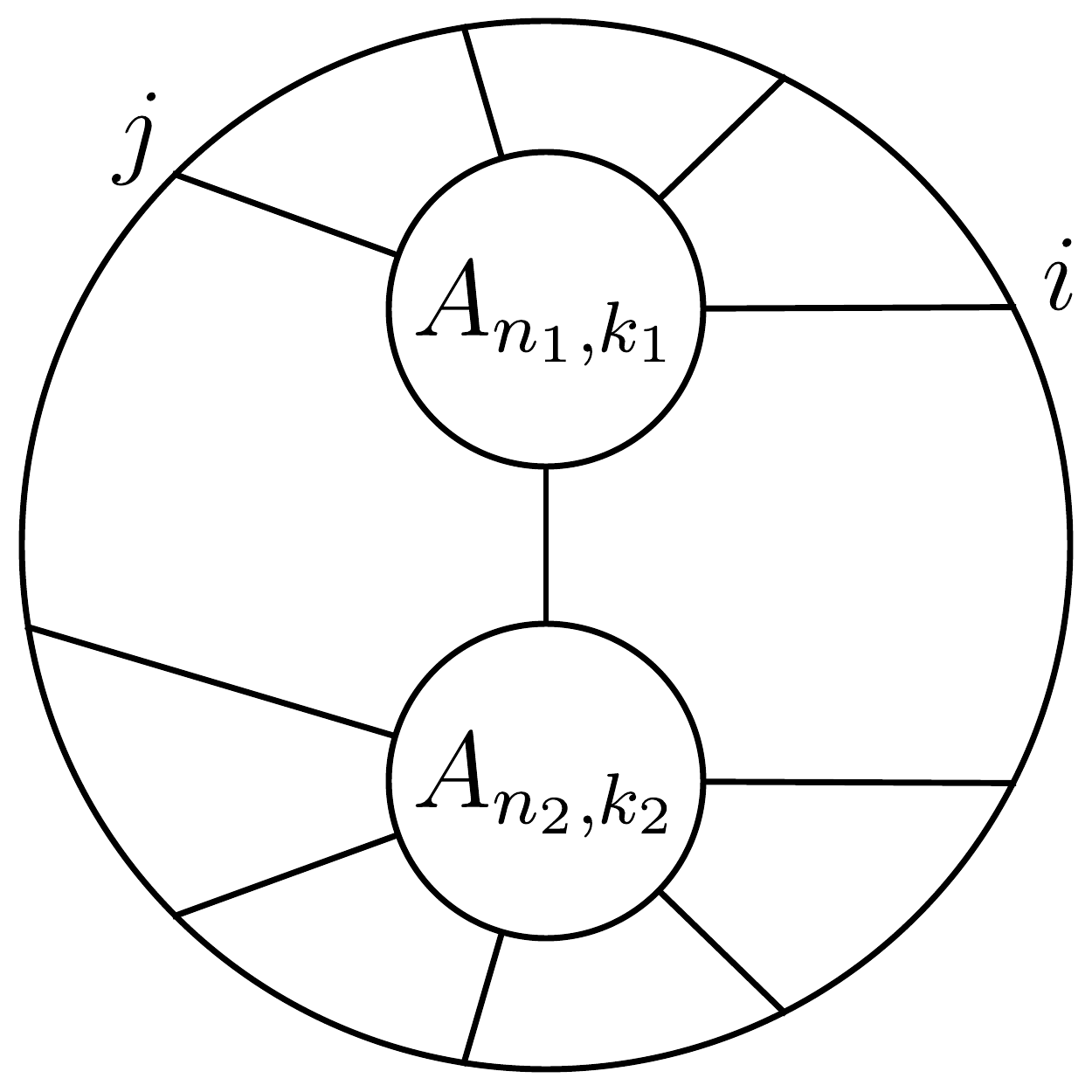}\\
$\langle Yii+1\rangle=0$&
$[ \tilde Yii+1]=0$&
$S_{ii+1\ldots j}=0$
\end{tabular}
\caption{Plabic diagrams for codimension one boundaries of the momentum amplituhedron.}
\label{fig:collinear}
\end{center}
\end{figure}

Let us take a more careful look at the structure of boundaries of $\mathcal{M}_{n,k}$. We have already specified that codimension one boundaries come in three different types: two of them are the collinear limits given by $\langle Y i i+1\rangle=0$ or $[\tilde Y i i+1]=0$, and there are factorization channels corresponding to $S_{i,i+1,\ldots,j}=0$. To each of these boundaries we can associate a plabic diagram which labels the corresponding positroid cell. They are of the form presented in Fig.~\ref{fig:collinear}. Notice that the two collinear limits can be understood as factorization channels with the top cell plabic diagram for three-particle amplitudes $A_{3,1}$ (white trivalent vertex) or $A_{3,2}$ (black trivalent vertex) attached as connected sub-diagrams.

\begin{figure}[h!]
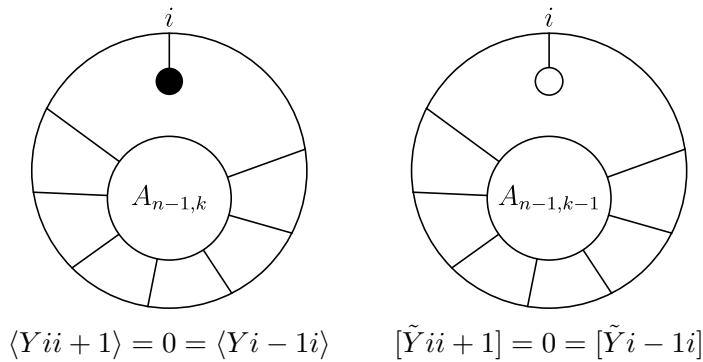

\begin{center}
\begin{tabular}{ccc}
\includegraphics[scale=0.3]{images/softwhite}&\quad&\includegraphics[scale=0.3]{images/softblack}\\
$\langle Yii+1\rangle=0=\langle Yi-1i\rangle$&&
$[ \tilde Yii+1]=0=[ \tilde Yi-1i]$
\end{tabular}
\caption{Plabic diagrams for codimension two boundaries of the momentum amplituhedron corresponding to physical soft limits of an amplitude.}
\label{fig:soft}
\end{center}
\end{figure}

\begin{figure}[h!]
\begin{center}
\includegraphics[scale=0.3]{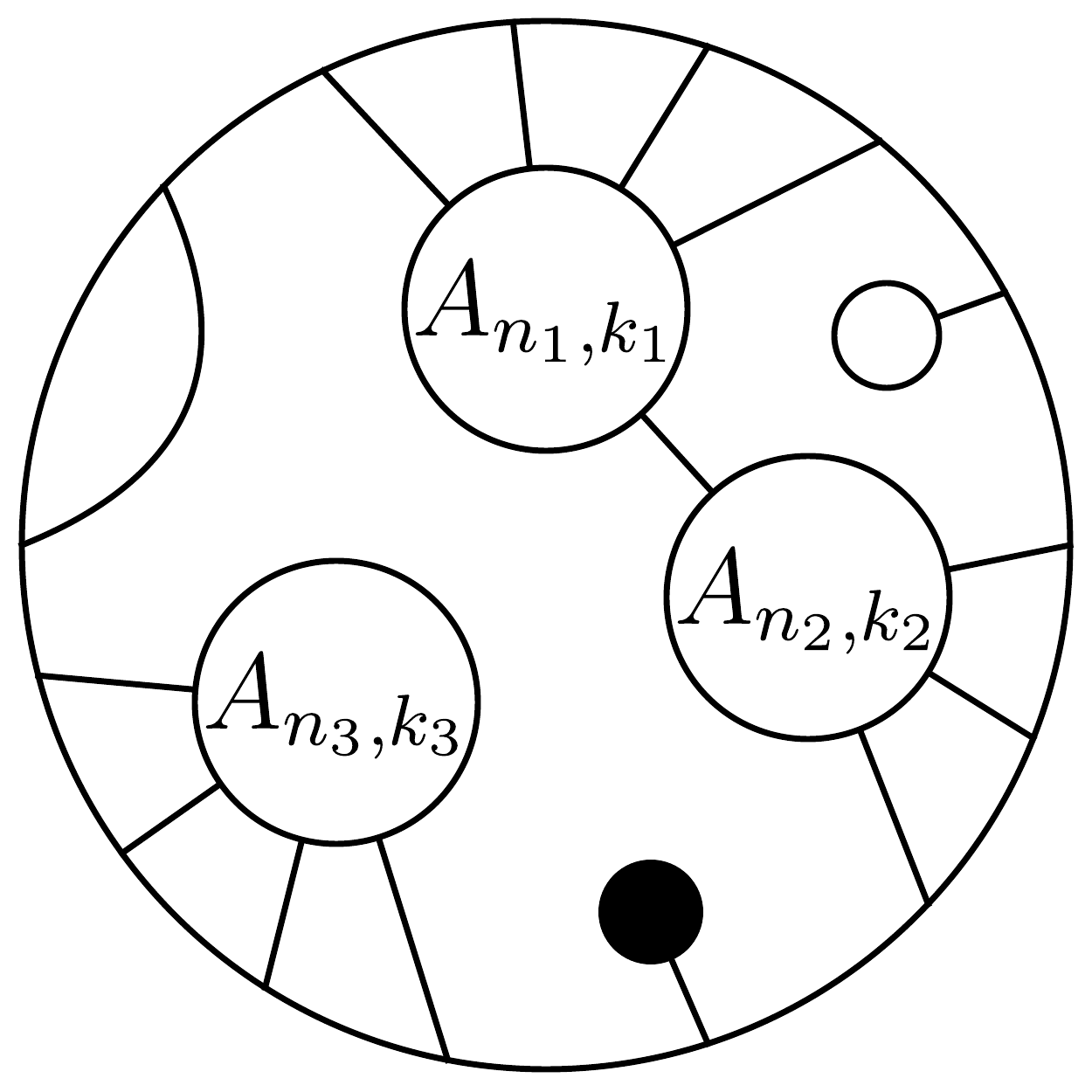}
\caption{Plabic diagram for a generic boundary of the momentum amplituhedron.}
\label{fig:genericboundary}
\end{center}
\end{figure}

At codimension two we find that the boundaries are either further factorizations of the factors in the previous step or further collinear limits. Concerning the latter, an interesting new behaviour emerges when we take the intersection of two consecutive codimension one boundaries corresponding to collinear limits of the same type. In this case, the plabic diagram corresponding to this boundary is a lollipop detached from the top cell diagram for a lower point amplitude, see Fig.~\ref{fig:soft}. These are exactly the soft limits which we described in the previous section.
These features can be generalized deeper into the geometry.
A generic boundary of the momentum amplituhedron will be a combination of collinear limits and factorizations, and the corresponding plabic diagram will have the generic form  depicted  in Fig.~\ref{fig:genericboundary}.
In particular, a plabic graph for a generic boundary of $\mathcal{M}_{n,k}$ will consist of a number of disjoint pieces, which can be of the following form:
\begin{itemize}
\item a black lollipop -- corresponding to a helicity-preserving soft limit
\item a white lollipop -- corresponding to a helicity-reducing soft limit
\item a single line -- corresponding to a forward-limit
\item a top cell, a collinear limit or a factorization channel for an amplitude $A_{n',k'}$ with $n'<n$ and $k'\le k$. In particular, it can be any boundary of $\mathcal{M}_{n',k'}$ as long as it is given by a connected diagram.
\end{itemize}
We can now use the package \texttt{amplituhedronBoundaries} to find all such boundaries of all dimensions from $d=0$ to $d=2n-4$. For $k=2$ and $k=n-2$ the boundaries of $\mathcal{M}_{n,2}$ and $\mathcal{M}_{n,n-2}$ trivially agree with the boundary stratifications of the positive Grassmannians $G_+(2,n)$ and $G_+(n-2,n)$, respectively, which are identical to each other via the Grassmannian duality. For $2<k<n-2$ the number of boundaries of a given dimension are organized in Table \ref{tab:fvectors}. Importantly, one can check that the Euler characteristic of the momentum amplituhedron $\mathcal{M}_{n,k}$ is equal to one for each of the cases we studied. We have also checked for the first few non-trivial cases that the poset of boundaries is Eulerian.

We conclude this section with the observation that there is an easy way to calculate $\dim_M \pi$, the momentum amplituhedron dimension of a given boundary, directly from its graphic representation. Consider first a connected diagram which depicts factorizations and collinear limits. Connected diagrams are always composed of lower-point amplitudes $(A_{n_1,k_1},A_{n_2,k_2},\ldots,A_{n_p,k_p})$ connected together into a tree graph. For this type of diagram we find that its dimension is given by
\begin{equation}
\left(\sum_{j=1}^p (2n_j-4)\right)-(p-1) \,,
\end{equation}
where the sum counts the dimension of the images of top cells for each of the lower-point amplitudes, and we subtract one dimension for each connecting internal edge in the tree. For the disconnected diagrams it is sufficient to add the dimensions of all disconnected pieces. Finally, every lollipop counts with dimension zero and every line with dimension one. This is demonstrated in the following example:
\begin{equation}
\dim_M\raisebox{-1cm}{\includegraphics[scale=0.2]{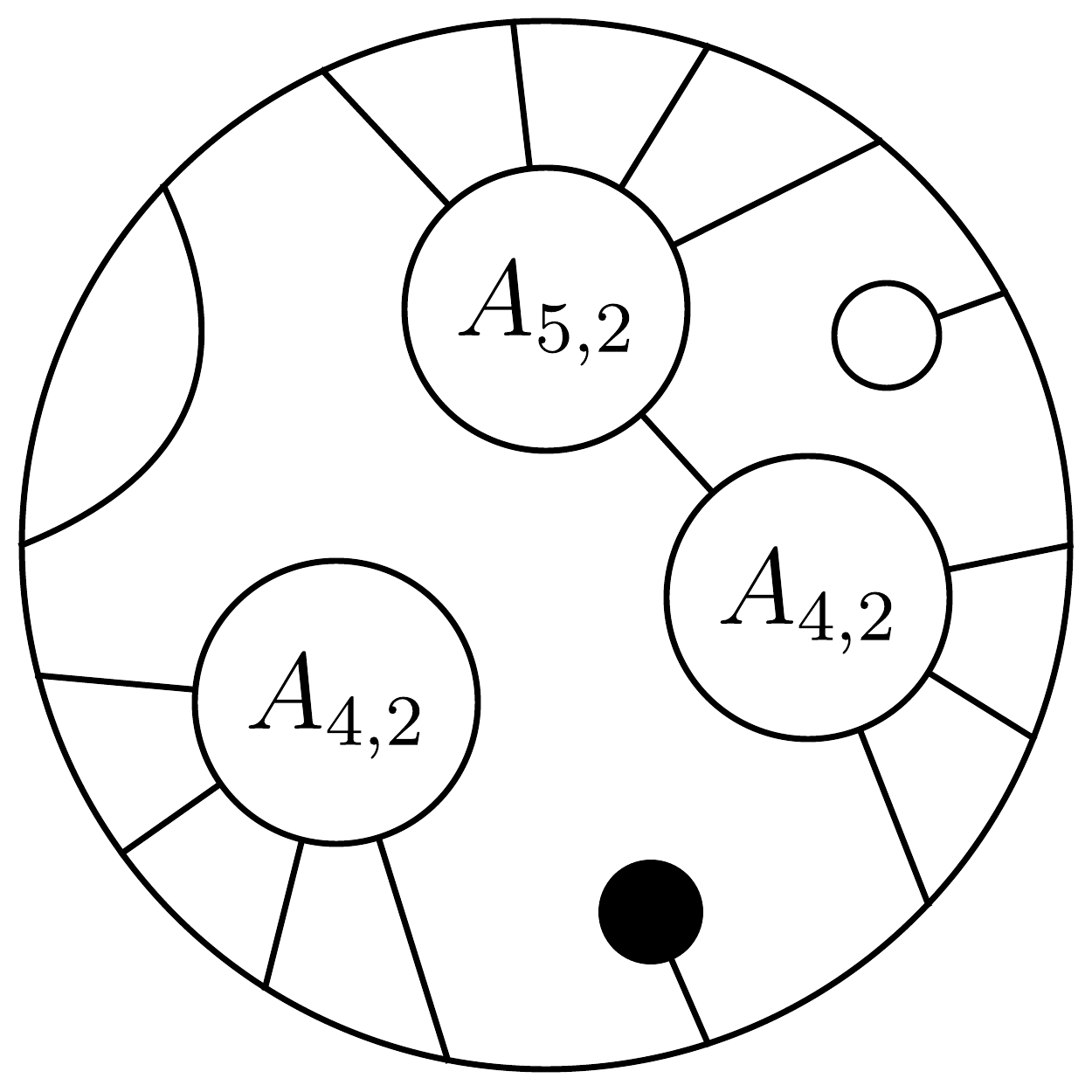}}=0+0+1+(2\cdot 5-4+2\cdot 4-4-1)+(2\cdot 4-4)=14 \,.
\end{equation}

%%%%%%%%%%%%%%%%%%%%%%%%%%%%%%%%%%

\subsection{Dual Graph Representation}
\label{sec:dual}

Having explicitly found all boundaries of the momentum amplituhedron for up to $n=9$, we now propose an efficient enumeration for these boundaries which enables us to extend our analysis beyond $n=9$ and to conjecture all boundaries for up to $n=11$. This enumeration is based on the notion of dual graphs where each boundary is labelled by a partial triangulation of a regular $n$-sided polygon (or $n$-gon) together with some additional decorations.

\begin{figure}
	\begin{center}
		\begin{tabular}{ccc}
			\includegraphics[width=0.25\textwidth]{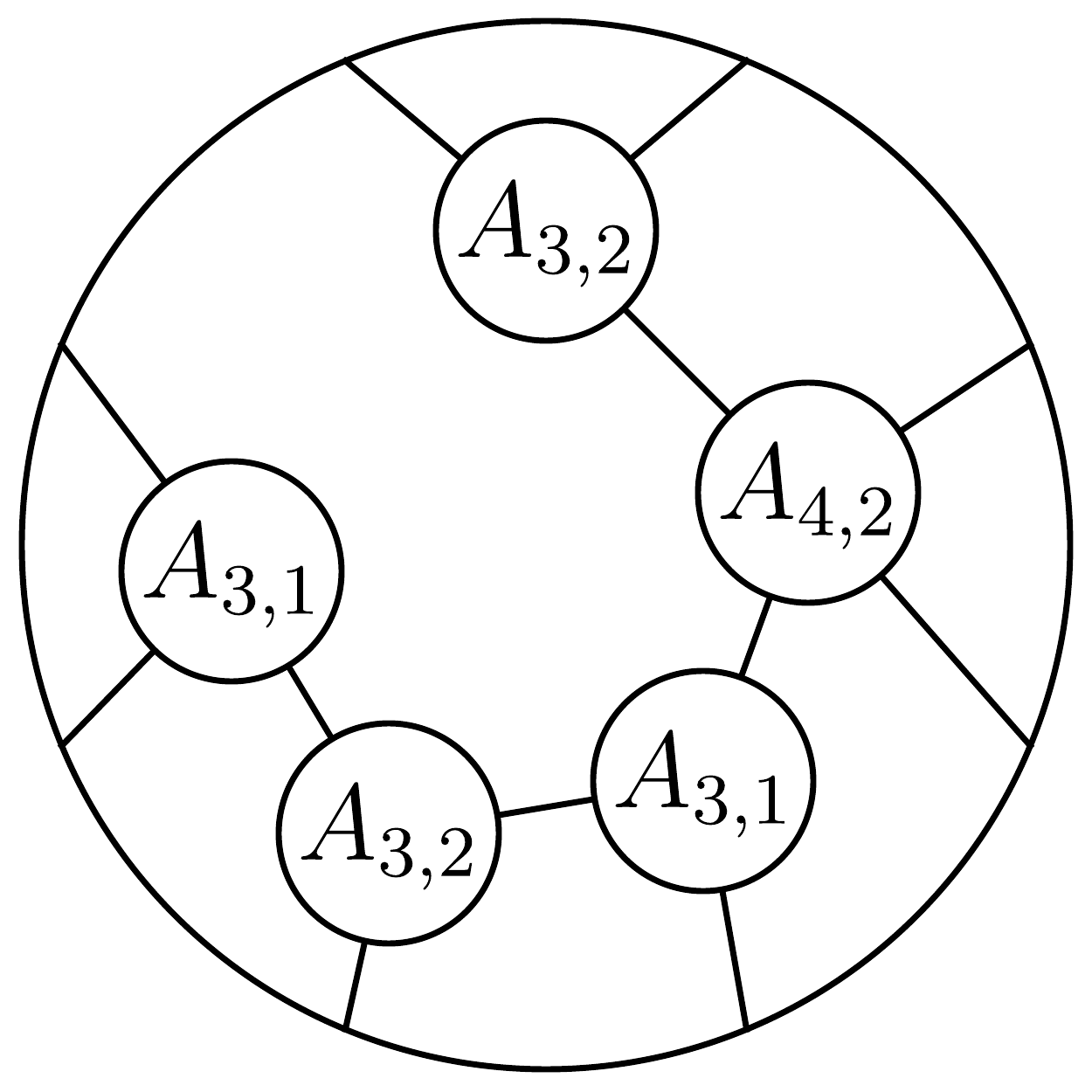}&\quad&
			\includegraphics[width=0.25\textwidth]{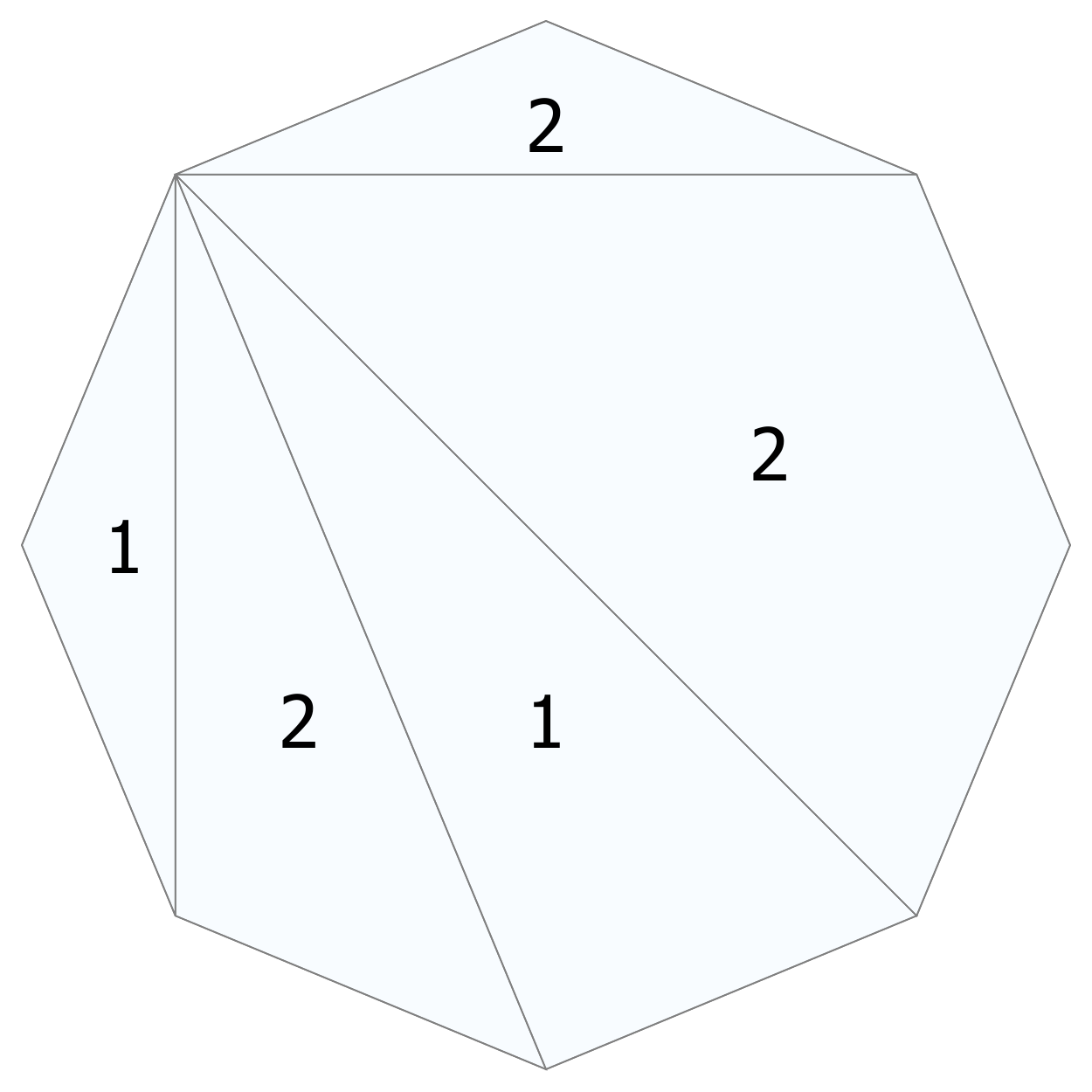}
		\end{tabular}
	\end{center}
	\caption{Dualization of a connected component of a boundary of the momentum amplituhedron.}
	\label{fig:dual1}
\end{figure}

Consider an on-shell diagram corresponding to an arbitrary boundary of the momentum amplituhedron. In general, such an on-shell diagram is a disconnected graph comprised of a collection of connected sub-diagrams. We shall focus first on dualizing the connected sub-diagrams and later reassemble them to obtain a diagrammatic label for the full on-shell diagram.
Each connected component is a tree consisting of $n'$ external legs and a finite number of internal vertices, where each vertex is an on-shell diagram corresponding to the top cell of some positive Grassmannian $G_+(k'',n'')$. For the present discussion, suppose $n'>1$. Dualizing such a tree graph, one obtains a labelled subdivision of an $n'$-gon as depicted in Fig.~\ref{fig:dual1}. 
For each element of the subdivision, we need to indicate which top cell on-shell diagram it represents, but since $n''$ is precisely the number of vertices of each such element, it is enough to specify the value of $k''$ only which we have indicated inside each element of the subdivision.
For each connected component, it is very easy to generate all possible subdivisions of an $n'$-gon and fill it in with all allowed $k''$ values. Once all such labelled subdivisions have been generated, there is one more thing to take into account: if in the connected sub-diagram (which is a tree graph) we have two $A_{3,1}$ (respectively\ $A_{3,2}$) amplitudes joined by an edge, then one can obtain an equivalent representative of the same sub-diagram using the so-called flip move. In order to address this possible overcounting, we need to quotient by this equivalence relation. 

\begin{figure}
	\begin{center}
		\begin{tabular}{ccc}
			\includegraphics[width=0.25\textwidth]{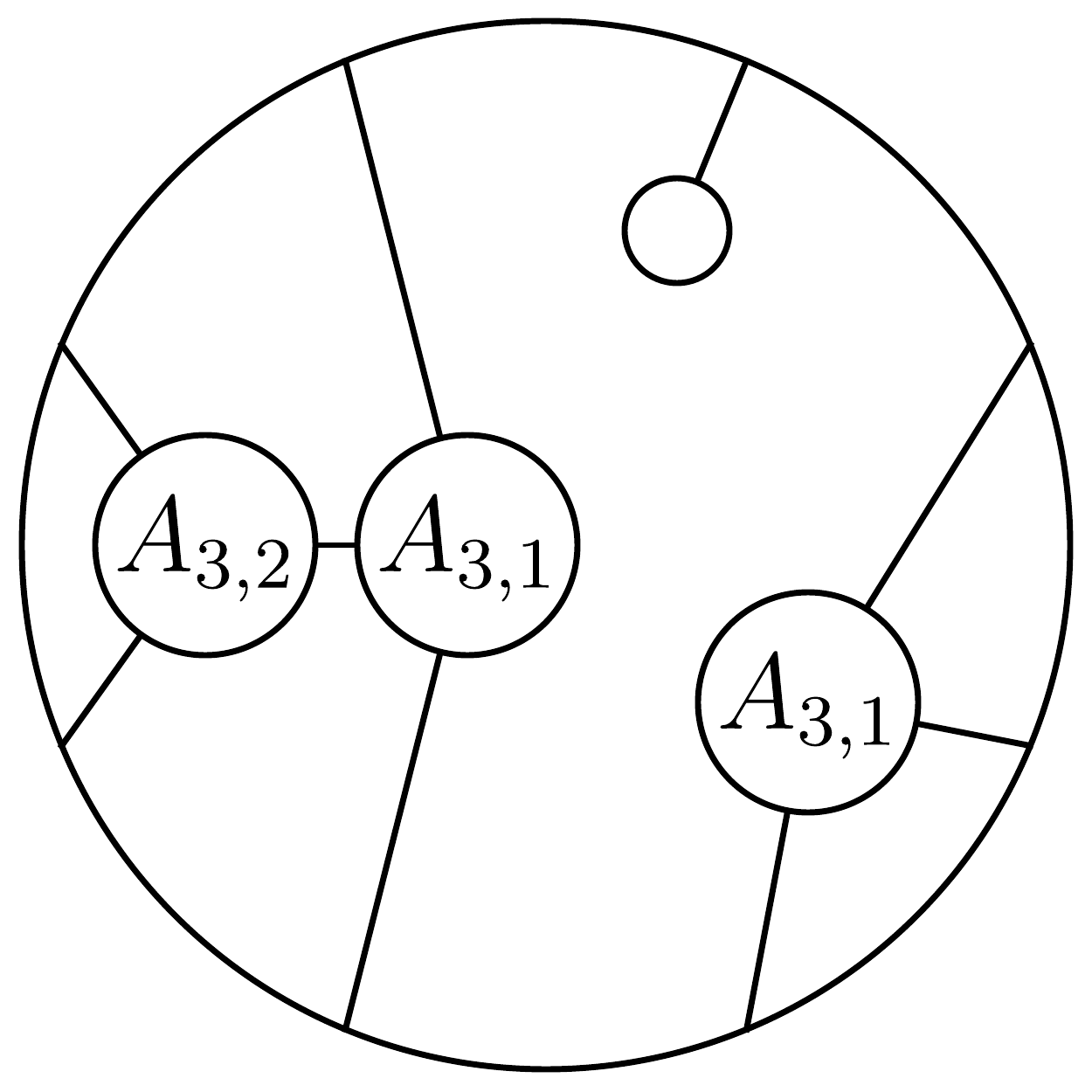}&\quad&
			\includegraphics[width=0.25\textwidth]{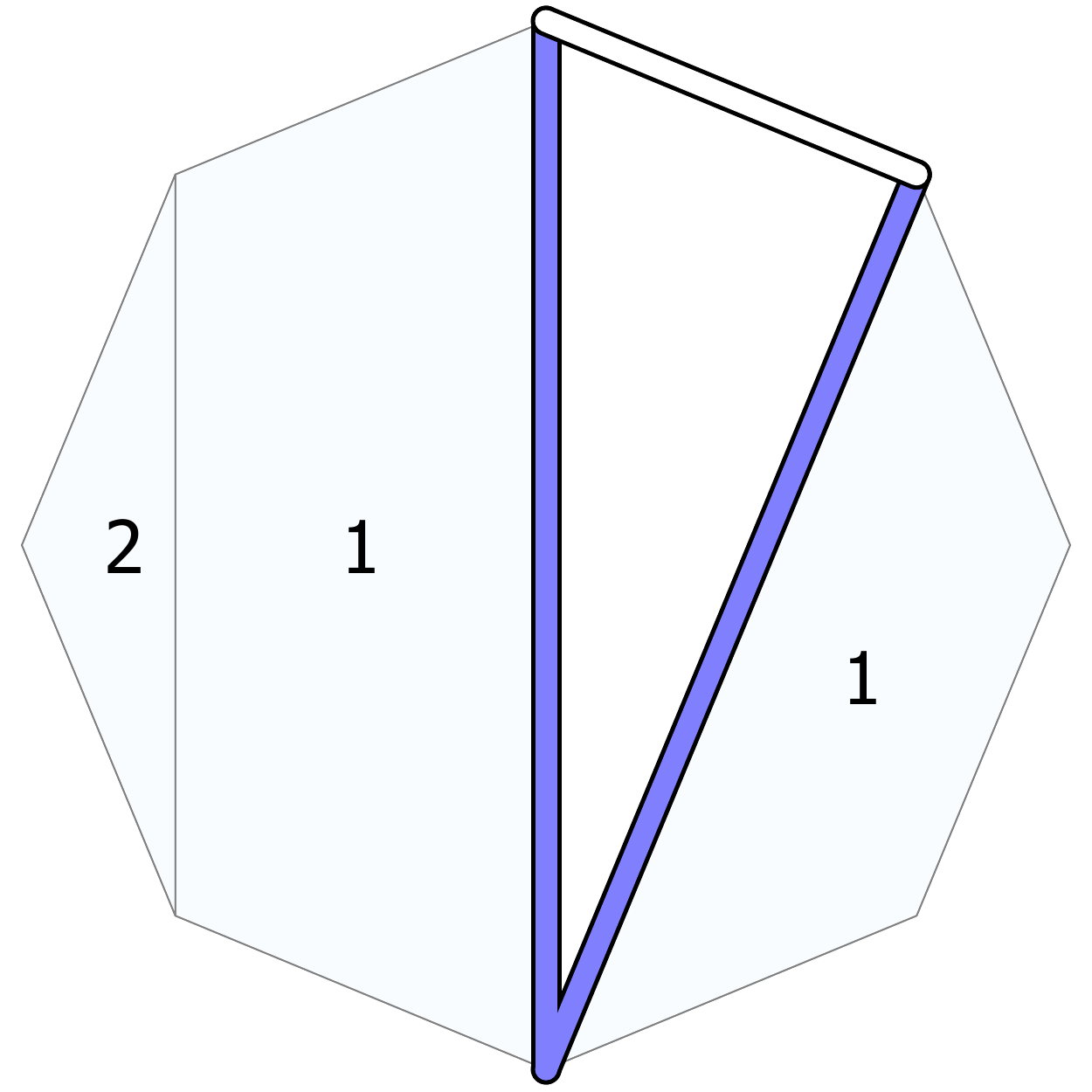}\\
			\includegraphics[width=0.25\textwidth]{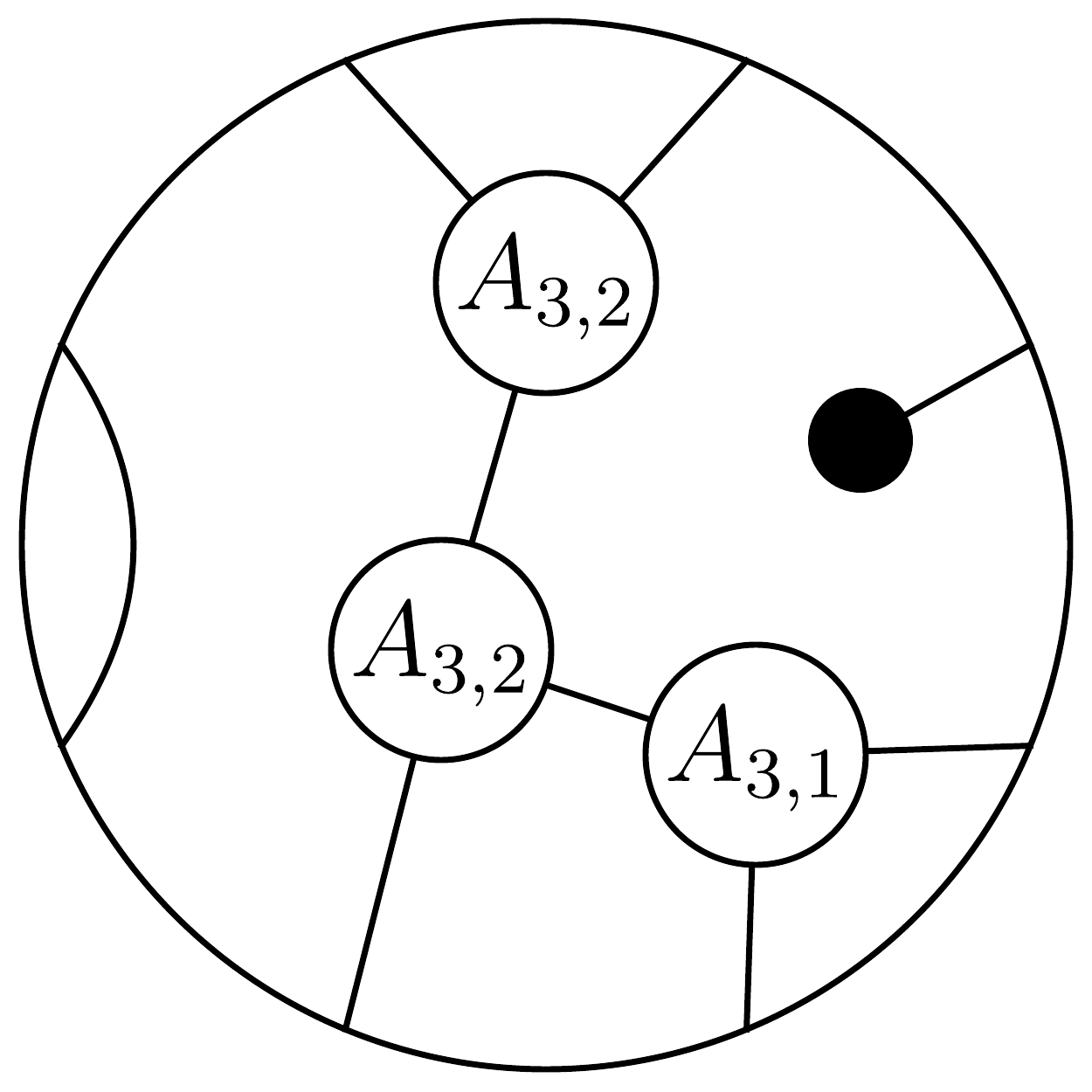}&\quad&
			\includegraphics[width=0.25\textwidth]{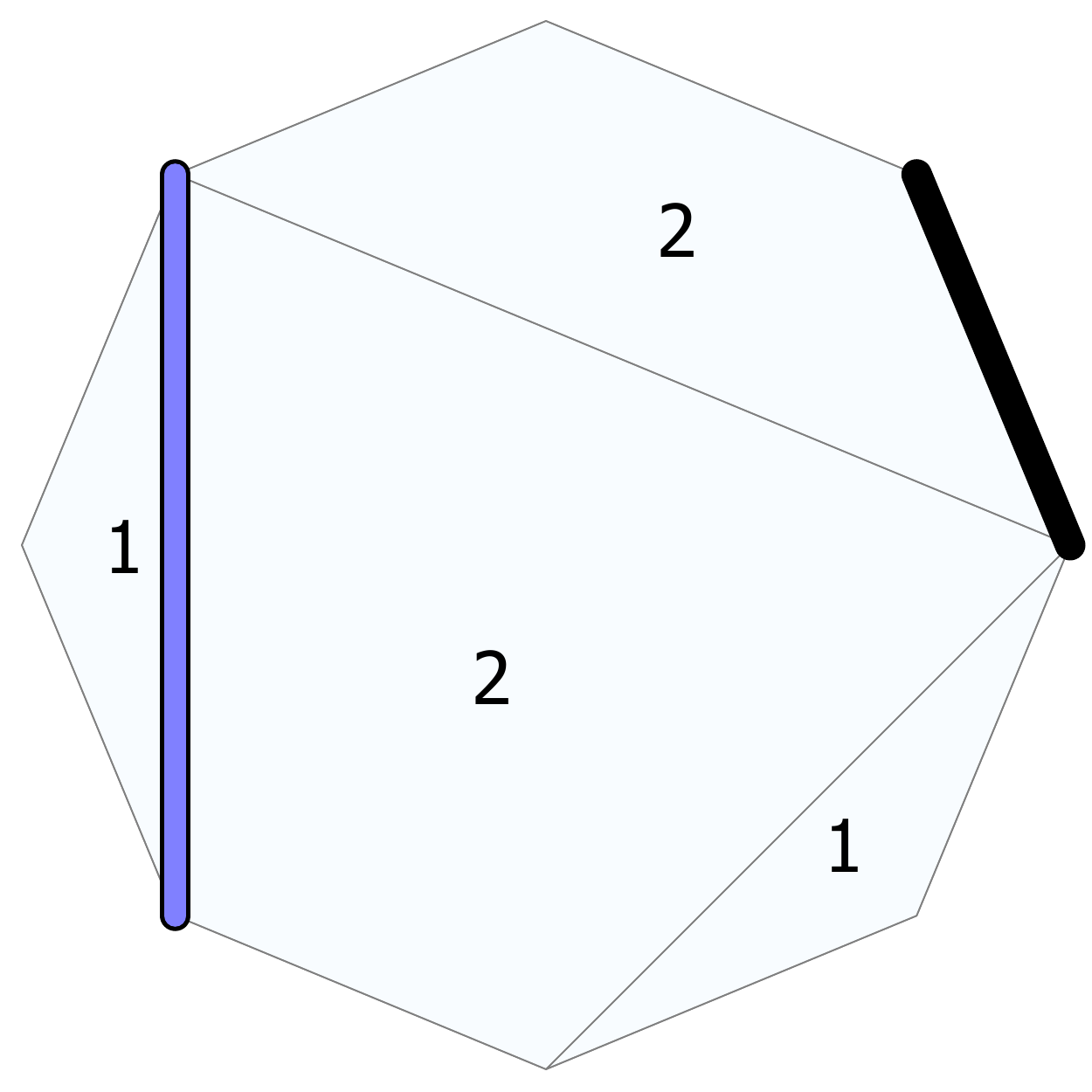}
		\end{tabular}
	\end{center}
	\caption{Dual diagrammatic labels for boundaries of the momentum amplituhedron.}
	\label{fig:dual2}
\end{figure}

Having generated all possible connected pieces, we then combine them in all possible ways, keeping in mind how the total helicity of an amplitude is related to the helicities of its connected pieces.
To each boundary, the combined diagrammatic label consists of the dual graphs for each connected component fitted together to form a partial triangulation of the $n$-gon as shown e.g.\ in Fig.~\ref{fig:dual2}. Borders between disconnected pieces are highlighted by thick internal edges. Finally, lollipop sub-diagrams, i.e.\ components with single external legs, are denoted by thick external edges. In particular, the white and black lollipops are depicted by thick white and black external edges, respectively. 

In this way we are able to (conjecturally) generate all boundaries for the momentum amplituhedron $\mathcal{M}_{n,k}$ for up to $n=11$ and for any $k$.

%%%%%%%%%%%%%%%%%%%%%%%%%%%%%%%%%%%%%%

%%%%%%%%%%%%%%%%%%%%%%%%%%%%%%%%%%%%%%%

\subsection{Generating Function for Boundaries}
Let us denote by $\mathcal{B}_{n,k}$ the set of all boundaries of the momentum amplituhedron $\mathcal{M}_{n,k}$. The set $\mathcal{B}_{n,k}$ is naturally divided into different sectors labelled by the momentum amplituhedron dimension. Therefore, knowing the number of momentum amplituhedron boundaries of all dimensions, we can construct the following generating function 
\begin{equation}
F_{n,k}(x)=\sum_{\sigma\in\mathcal{B}_{n,k}}(-x)^{\dim_M \sigma} \,.
\end{equation}
For $n\leq 9$ this function can be easily found by using the data available in Table \ref{tab:fvectors}. We provide the explicit forms of the generating functions $F_{n,k}$ for $9\leq n\leq 11$ in Table \ref{table:generating.high}.
In particular, in all the cases we studied we found that
\begin{equation}
F_{n,k}(1)=1 \,,
\end{equation}
which implies that the Euler characteristic of the momentum amplituhedron equals 1.

Presently we do not know the general form of the generating function $F_{n,k}$ for arbitrary $n$ and $k$. It is, however, worthwhile to note that a corresponding generating function for the positive Grassmannian $G_+(k,n)$ has been found in \cite{williams2003enumeration}. The method used there relied on finding a recursion relation for the number of $\Le$-diagrams \cite{Postnikov:2006kva} of a given type. It remains an open question whether a similar calculation can be repeated for the momentum amplituhedron boundaries which are described by a subset of these $\Le$-diagrams. As a first step in this direction, we classify the $\Le$-diagrams corresponding to permutations in the positive Grassmannian which do not label momentum amplituhedron boundaries for $k=3$ and $k=4$. Consider the first non-trivial\footnote{$\mathcal{M}_{6,3}$ is the first example whose boundary stratification is not isomorphic to that of the positive Grassmannian $G_+(3,6)$.} momentum amplituhedron example: $\mathcal{M}_{6,3}$. The cells in the positive Grassmannian $G_+(3,6)$ which are not in $\mathcal{M}_{6,3}$ are labelled by the following $\Le$-diagrams:
\begin{align}
\vcenter{\hbox{\includegraphics[scale=0.8]{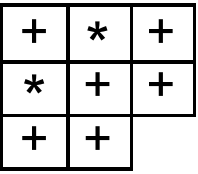}}}~,~
\vcenter{\hbox{\includegraphics[scale=0.8]{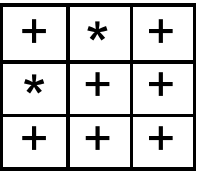}}}~,~
\vcenter{\hbox{\includegraphics[scale=0.8]{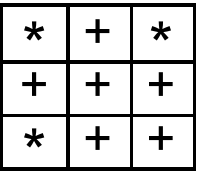}}}~,\label{eq:badLeDiagrams-k=3}
\end{align}
where $\ast$ can be either $+$ or $0$, excluding the case when the last two diagrams are completely populated with $+$ symbols as this labels the top cell of $G_+(3,6)$.
For $k=3$, we conjecture that for all $n$, a cell in $G_+(3,n)$ is not in $\mathcal{M}_{n,3}$ if and only if its associated $\Le$-diagram contains one of the $\Le$-diagrams in \eqref{eq:badLeDiagrams-k=3} as a sub-diagram. We have explicitly confirmed this result for $n=7,8,9$. For $k=4$, the above criterion is not sufficient, and one needs to include additional ``bad'' $\Le$-diagrams. In particular, for $G_+(4,8)$ we find that the above condition captures all but one cell whose $\Le$-diagram is given by
\begin{align}
\vcenter{\hbox{
\includegraphics[scale=0.8]{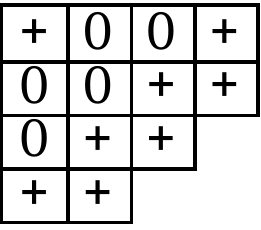}}}~.
\label{eq:badLeDiagrams-8,4}
\end{align}
Interestingly, we find that all cells $G_+(4,9)$ which are not in $\mathcal{M}_{9,4} $ are labelled by $\Le$-diagrams which contain either one of the $\Le$-diagrams in \eqref{eq:badLeDiagrams-k=3} or the one in \eqref{eq:badLeDiagrams-8,4} as a sub diagram.

%%%%%%%%%%%%%%%%%%%%%%%%%%%%%%%%%%%%%
%%%%%%%%%%%%%%%%%%%%%%%%%%%%%%%%%%%%%%

\section{Conclusions and Outlook}
In this paper we classified all physical singularities of tree-level scattering amplitudes in $\mathcal{N}=4 $ sYM by studying the boundaries of the momentum amplituhedron $\mathcal{M}_{n,k}$. Each singularity comes from a subsequent multi-particle factorization and collinear limit of the amplitude, which can be translated to geometry as an appropriate intersection of facets of the momentum amplituhedron. 
There are a few natural questions to investigate following our work. From a mathematical point of view, we have laid the foundation for proving that the momentum  amplituhedron $\mathcal{M}_{n,k}$ is a ball. In particular, we showed that, for all the cases we analysed, the Euler characteristic is equal to one. It would also be desirable to find a general form of the generating function, similar to the one for the positive Grassmannian found in \cite{williams2003enumeration}, to show that this feature holds for any $n$ and $k$. Moreover, we showed that in the first few cases the momentum amplituhedron has an Eulerian boundary poset. The fact that the momentum amplituhedron has the combinatorial structure of a ball provides evidence that it is a simpler geometry compared to the ordinary amplituhedron, for which our preliminary studies indicate that the boundary structure is more complicated. 
From the point of view of physics, the natural question to ask is whether we can extend this analysis beyond tree level. At the moment, the loop-level momentum amplituhedron, i.e.\ the geometry associated to scattering amplitudes at loop level in spinor-helicity space, is not known. Once its definition is found, the methods from this paper suggest a natural starting point for finding and studying its boundary structure.

\section{Acknowledgements}
This work was partially funded by the Deutsche Forschungsgemeinschaft (DFG, German Research Foundation) -- Projektnummern 404358295 and 404362017.

\appendix
%\section{Generating Functions}
\begin{table}
\begin{align*}
%F_{4,2}&=x^4-4 x^3+10 x^2-12 x+6\\
%F_{5,2}&=x^6-5 x^5+15 x^4-30 x^3+40 x^2-30 x+10\\
%F_{6,2}&=x^8-6 x^7+21 x^6-50 x^5+90 x^4-120 x^3+110 x^2-60 x+15\\
%F_{6,3}&=x^8-15 x^7+54 x^6-114 x^5+180 x^4-215 x^3+180 x^2-90 x+20\\
%F_{7,2}&=x^{10}-7 x^9+28 x^8-77 x^7+161 x^6-266 x^5+350 x^4-350 x^3+245 x^2-105 x+21\\
%F_{7,3}&=x^{10}-21 x^9+119 x^8-350 x^7+665 x^6-938 x^5+1050 x^4-910 x^3+560 x^2-210 x+35\\
%F_{8,2}&=x^{12}-8 x^{11}+36 x^{10}-112 x^9+266 x^8-504 x^7+784 x^6-1008 x^5+1050 x^4-840 x^3\\
%&+476 x^2-168 x+28\\
%F_{8,3}&=x^{12}-28 x^{11}+188 x^{10}-720 x^9+1820 x^8-3262 x^7+4424 x^6-4788 x^5+4200 x^4\\
%&-2870 x^3+1400 x^2-420 x+56\\
%F_{8,4}&=x^{12}-32 x^{11}+300 x^{10}-1280 x^9+3264 x^8-5696 x^7+7420 x^6-7672 x^5+6426 x^4\\
%&-4200 x^3+1960 x^2-560 x+70\\
F_{9,2}&=x^{14}-9 x^{13}+45 x^{12}-156 x^{11}+414 x^{10}-882 x^9+1554 x^8-2304 x^7+2898 x^6-3066 x^5\\
&+2646 x^4-1764 x^3+840 x^2-252 x+36\\
F_{9,3}&=x^{14}-36 x^{13}+279 x^{12}-1227 x^{11}+3726 x^{10}-8370 x^9+14322 x^8-19152 x^7+20622 x^6\\
&-18270 x^5+13230 x^4-7476 x^3+3024 x^2-756 x+84\\
F_{9,4}&=x^{14}-45 x^{13}+540 x^{12}-3003 x^{11}+10089 x^{10}-23049 x^9+38298 x^8-48618 x^7\\
&+49140 x^6-40656 x^5+27468 x^4-14490 x^3+5460 x^2-1260 x+126\\
F_{10,2}&=x^{16}-10 x^{15}+55 x^{14}-210 x^{13}+615 x^{12}-1452 x^{11}+2850 x^{10}-4740 x^9+6765 x^8\\
&-8340 x^7+8862 x^6-7980 x^5+5880 x^4-3360 x^3+1380 x^2-360 x+45\\
F_{10,3}&=x^{16}-45 x^{15}+395 x^{14}-1955 x^{13}+6705 x^{12}-17412 x^{11}+35640 x^{10}-58440 x^9\\
&+77490 x^8-84120 x^7+75852 x^6-57120 x^5+35280 x^4-17010 x^3+5880 x^2\\
&-1260 x+120\\
F_{10,4}&=x^{16}-60 x^{15}+880 x^{14}-5780 x^{13}+23385 x^{12}-65990 x^{11}+137835 x^{10}-220662 x^9\\&
+277890 x^8-281940 x^7+235410 x^6-163380 x^5+92862 x^4-41160 x^3+13020 x^2\\
&-2520 x+210\\
F_{10,5}&=x^{16}-65 x^{15}+1045 x^{14}-7915 x^{13}+34740 x^{12}-101240 x^{11}+212285 x^{10}-336220 x^9\\
&+415890 x^8-412980 x^7+336840 x^6-228102 x^5+126420 x^4-54600 x^3+16800 x^2\\
&-3150 x+252\\
F_{11,2}&=x^{18}-11 x^{17}+66 x^{16}-275 x^{15}+880 x^{14}-2277 x^{13}+4917 x^{12}-9042 x^{11}+14355 x^{10}\\
&-19855 x^9+24057 x^8-25542 x^7+23562 x^6-18480 x^5+11880 x^4-5940 x^3+2145 x^2\\
&-495 x+55\\
F_{11,3}&=x^{18}-55 x^{17}+539 x^{16}-2959 x^{15}+11275 x^{14}-32692 x^{13}+75735 x^{12}-143913 x^{11}\\
&+226908 x^{10}-297990 x^9+326997 x^8-301620 x^7+235158 x^6-154308 x^5+83160 x^4\\
&-34980 x^3+10560 x^2-1980
   x+165\\
   F_{11,4}&=x^{18}-77 x^{17}+1342 x^{16}-10109 x^{15}+46849 x^{14}-153527 x^{13}+380402 x^{12}-738067 x^{11}\\
   &+1143780 x^{10}-1435005 x^9+1475562 x^8-1259412 x^7+901362 x^6-540540 x^5\\
   &+265650 x^4-101640 x^3+27720
   x^2-4620 x+330\\
   F_{11,5}&=x^{18}-88 x^{17}+1782 x^{16}-16522 x^{15}+88924 x^{14}-318197 x^{13}+820512 x^{12}\\
   &-1602986 x^{11}+2450437 x^{10}-2996972 x^9+2984520 x^8-2457840 x^7+1693230 x^6
   \\&-975744 x^5+460152 x^4-168630 x^3+43890
   x^2-6930 x+462\\
   F_{12,2}&=x^{20}-12 x^{19}+78 x^{18}-352 x^{17}+1221 x^{16}-3432 x^{15}+8074 x^{14}-16236 x^{13}+28314 x^{12}\\
   &-43252 x^{11}+58278 x^{10}-69564 x^9+73656 x^8-68904 x^7+56232 x^6-39072 x^5\\
   &+22275 x^4-9900 x^3+3190
   x^2-660 x+66
\end{align*}
\caption{Generating functions for momentum amplituhedron boundaries. Recall that $F_{n,k}=F_{n,n-k}$}
\label{table:generating.high}
\end{table}

\pagebreak

\bibliographystyle{nb}

\bibliography{mom_ampl}

\end{document}